\documentclass[12pt]{article}

\usepackage{jheppub}

\usepackage{latexsym,amsmath,amsfonts,amssymb}
\usepackage{mathrsfs}
\usepackage[makeroom]{cancel}
\usepackage{bbm}
\usepackage{bm}
\usepackage{subfigure}
\usepackage{paralist}
\usepackage{youngtab}

\newcommand{\vev}[1]{\langle #1 \rangle}

\numberwithin{equation}{section}

\newcommand{\nn}{\nonumber}

\newcommand{\be}{\begin{equation}} 
\newcommand{\ee}{\end{equation}}
\newcommand{\bea}{\begin{equation} \begin{aligned}} \newcommand{\eea}{\end{aligned} \end{equation}}

\newcommand{\bit}{\begin{itemize}} 
\newcommand{\eit}{\end{itemize}}

\newcommand{\cH}{\mathcal{H}}

\newcommand{\cO}{\mathcal{O}}

\newcommand{\cS}{\mathcal{S}}

\newcommand{\bZ}{\mathbb{Z}}

\newcommand{\Z}{\mathbb{Z}}
\newcommand{\C}{\mathbb{C}}

\renewcommand{\t}{\widetilde }
\renewcommand{\d}{\partial }

\renewcommand{\b}{\bar }

\newcommand{\half}{{1\over 2}}

\newcommand{\bz}{{\b z}}

\newcommand{\CH}{\mathcal{H}}

\newcommand{\CM}{\mathcal{M}}
\newcommand{\CN}{\mathcal{N}}
\newcommand{\CO}{\mathcal{O}}

\newcommand{\CS}{\mathcal{S}}
\newcommand{\CT}{\mathcal{T}}

\newcommand{\CV}{\mathcal{V}}
\newcommand{\CW}{\mathcal{W}}

\newcommand{\CZ}{\mathcal{Z}}

\newcommand{\FR}{\mathfrak{R}}
\newcommand{\Fg}{\mathfrak{g}}
\newcommand{\Fh}{\mathfrak{h}}

\newcommand{\GG}{\mathbf{G}}

\newcommand{\GH}{\mathbf{H}}

\newcommand{\rk}{{{\rm rk}(\GG)}}

\newcommand{\m}{\mathfrak{m}}
\newcommand{\n}{\mathfrak{n}}
\newcommand{\p}{\mathfrak{p}}

\newcommand{\ba}{{\b a}}
\newcommand{\bb}{{\b b}}

\newcommand{\h}{\hat}

\newcommand{\s}{\sigma}

\newcommand{\pif}{{\Pi}}

\DeclareMathOperator{\Tr}{Tr}
\DeclareMathOperator{\tr}{tr}

\newcommand{\SL}{{\mathscr L}}

\newcommand{\eq}[1]{(\ref{#1})}
\newcommand{\ov}{\over}

\newcommand{\fM}{\mathfrak{M}}

\newcommand{\kay}{k}
\newcommand{\sig}{\sigma}

\newcommand{\ea}{\end{array}}
\newcommand{\bi}{\begin{itemize}}
\newcommand{\ei}{\end{itemize}}
\def\vec#1{\bm{#1}}
 \newcommand{\ben}{\begin{enumerate}}
\newcommand{\een}{\end{enumerate}}
\newcommand{\bean}{\begin{eqnarray*}}
\newcommand{\eean}{\end{eqnarray*}}
\newcommand{\eref}[1]{(\ref{#1})}

\newcommand{\BP}{\mathbb{P}}

\newcommand{\comment}[1]{}

\preprint{CERN-TH-2017-107}

\title{$A$-twisted correlators and Hori dualities}

\author[1]{Cyril Closset,}
\author[2,3]{Noppadol Mekareeya,}
\author[4]{Daniel S. Park}

\affiliation[1]{Theory Department, CERN\\
CH-1211, Geneva 23, Switzerland}

\affiliation[2]{Dipartimento di Fisica, Universit\`a di Milano-Bicocca, \\ Piazza della Scienza 3, I-20126 Milano, Italy}
\affiliation[3]{INFN, sezione di Milano-Bicocca, I-20126 Milano, Italy}

\affiliation[4]{ NHETC and Department of Physics and Astronomy\\
Rutgers University, Piscataway, NJ 08855-0849, USA}

\abstract{The Hori-Tong and Hori dualities are infrared dualities between two-dimensional gauge theories with $\CN=(2,2)$ supersymmetry, which are reminiscent of four-dimensional Seiberg dualities.  
We provide additional evidence for those dualities with $U(N_c)$, $USp(2N_c)$, $SO(N)$ and $O(N)$ gauge groups, by matching correlation functions of Coulomb branch operators on a Riemann surface  $\Sigma_g$, in the presence of the topological $A$-twist. The $O(N)$ theories studied, denoted by $O_+ (N)$ and $O_- (N)$, can be understood as $\mathbb{Z}_2$ orbifolds of an $SO(N)$ theory. The correlators of these theories on $\Sigma_g$ with $g  > 0$ are obtained by computing correlators with $\mathbb{Z}_2$-twisted boundary conditions and summing them up with weights determined by the orbifold projection.
}


\begin{document}

\setcounter{tocdepth}{2}

\maketitle



\section{Introduction}
Two-dimensional supersymmetric gauge theories have a rich dynamics,  similar to the one of their higher-dimensional cousins. In particular, two-dimensional gauge theories with $\CN=(2,2)$ supersymmetry admit infrared-dual descriptions \cite{Hori:2006dk, Hori:2011pd, Benini:2012ui} reminiscent of four-dimensional $\CN=1$ Seiberg duality \cite{Seiberg:1994pq}.  Thanks to the renewal of supersymmetric localization techniques in two dimensions \cite{Benini:2012ui, Doroud:2012xw, Benini:2013nda, Benini:2013xpa}---see \cite{Benini:2016qnm, 2DReview} for recent reviews---one can provide highly non-trivial tests of infrared dualities by matching supersymmetric partition functions of dual theories. 
In addition, new exact expressions were also obtained for correlation functions of certain half-BPS local operators in two-dimensional non-abelian gauge theories \cite{Benini:2015noa, Closset:2015rna, Closset:2015ohf}, generalizing the seminal results of \cite{Witten:1993yc, Witten:1993xi, Morrison:1994fr}. See also \cite{Melnikov:2005hq, Melnikov:2005tk, Orlando:2010uu, Jia:2014ffa, Benini:2014mia, Gomis:2014eya,  Yamazaki:2015voa, Gomis:2015yaa, Bae:2015eoa, Guo:2015caf, Ueda:2016wfa, Gerhardus:2016iot, Cho:2017bhd} for related works. 

In this note, we study the matching of twisted chiral ring correlation functions across Seiberg-like dualities. Consider a gauge group $\GG$ of rank $N_c$, with  Lie algebra $\Fg$. We consider the ultraviolet-free, SQCD-like theory consisting of a $\Fg$-valued vector multiplet coupled to $N_f$ fundamental flavors---chiral multiplets in the fundamental representation of $\Fg$. Schematically, the ``electric'' and ``magnetic'' dual gauge groups are \cite{Hori:2006dk, Hori:2011pd, Benini:2012ui}:
\bea\label{sd gauge groups}
&U(N_c) \quad &\leftrightarrow  &&\quad & U(N_f-N_c)~, \cr
&USp(2N_c) \quad &\leftrightarrow  &&\quad & USp(N_f-2N_c-1)~, \cr
&SO(N) \quad &\leftrightarrow  &&\quad & O_+(N_f-N+1)~\cr
&O_-(N) \quad &\leftrightarrow  &&\quad & O_-(N_f-N+1)~.
\eea
Note that $N_f$ should be odd in $USp(2N_c)$ case. In the case of $SO(N)$, we can have $N= 2 N_c$ or $N=2 N_c+1$ while $N_f$ can be even or odd. In addition, there are distinct ways to define the action of the discrete $\Z_2$ in the $O(N)$ gauge group (the ``$\Z_2$ orbifold''), denoted $O_\pm(N)$. This  leads to a rich pattern of dualities, which were carefully studied by Hori  in \cite{Hori:2011pd}.
All the ``magnetic'' theories also contain ``mesons''---gauge singlet chiral multiplets $M$, which are coupled to the dual flavors through the superpotential. All the dualities are between so-called ``regular'' theories, which are theories without a quantum Coulomb branch \cite{Hori:2011pd}.

These two-dimensional theories have interesting `Coulomb branch' operators $\CO(\sigma)$, which are gauge-invariant polynomials in the $\Fg$-valued complex scalar field $\sigma$ that sits in the $\CN=(2,2)$ vector multiplet. For a $U(N_c)$ gauge group, for instance, we have:
\be
\CO_n = \Tr(\sigma^n)~, \qquad n=0,  1, \cdots, N_c~.
\ee
In simple-enough cases, like the ones we will consider, these operators generate the full {\it twisted chiral ring} of the theory.
We can compute their correlation functions {\it exactly} (including all instanton corrections) on a curved-space background preserving the two supercharges $Q_-$ and $\t Q_+$ that commute with $\sigma$, thanks to the topological $A$-twist \cite{Witten:1988xj, Witten:1993yc, Morrison:1994fr}. Let us consider  $\Sigma_g$ a closed Riemann surface of genus $g$.  The coupling of the field theory to the metric (and its superpartners) on $\Sigma_g$ depends on a choice of $R$-charges for the vector-like $U(1)_R$ symmetry. To preserve supersymmetry, we must have a flux
\be\label{flux R}
{1\ov 2 \pi} \int_{\Sigma_g} d A^{(R)} = g-1
\ee
for the $U(1)_R$ background gauge field $A_\mu^{(R)}$. This leads to the Dirac quantization condition  
\be\label{dirac quan R}
r (g-1) \in \Z~,
\ee 
with $r$ the $R$-charge \cite{Witten:1988xj, Closset:2014pda}.  In the presence of flavor symmetries (that is, any non-$R$ global symmetry), we may also turn on fluxes 
\be\label{def nF}
{1\ov 2 \pi} \int_{\Sigma_g} d A^{(F)} = \n_F \, \in \Z
\ee
for background gauge fields coupling to the conserved currents. (Naturally, $A_\mu^{(F)}$ sits in a  background vector multiplet $\CV^{(F)}$.) The correlation functions of Coulomb branch operators on $\Sigma_g$, with background fluxes \eqref{def nF} turned on, are given by \cite{Melnikov:2005tk, Nekrasov:2014xaa, Benini:2015noa, Closset:2015rna, Benini:2016hjo, Closset:2016arn}:
\be\label{gen formula correlators}
\left\langle \CO(\sigma) \right\rangle_{g; \,\n_F}  = \sum_{\h \sigma\in \CS_{\rm BE}} \CO(\h \sigma) \,\CH(\h\sigma)^{g-1} \, \prod_F \pif_{F}(\h\sigma)^{\n_F}~,
\ee
with $F$ an index running over the flavor group. The operator $\CH$ is the {\it handle-gluing operator}  \cite{Nekrasov:2014xaa} and  $\pif_F$ are  flavor {\it flux operators}  \cite{Closset:2017zgf}, as we will review. Those operators  are functions of $\sigma$, and the sum in \eqref{gen formula correlators} is over the distinct solutions $\sigma=\h\sigma$ of the associated {\it Bethe equations} \cite{Nekrasov:2009uh}---the saddle points of the Coulomb-branch effective twisted superpotential.

In this note, we study these correlation functions in two-dimensional SQCD-like theories and we prove the equality:
\be\label{dual rel}
\left\langle \CO(\sigma) \right\rangle_{g; \n}^{\CT}  = \left\langle \CO_D(\sigma_D) \right\rangle_{g; \n}^{\CT_D}
\ee
for any two theories $\CT$ and $\CT_D$ related by Hori duality as in \eqref{sd gauge groups}.  
This provides additional evidence for the dualities.
It is also an interesting application of the formula \eqref{gen formula correlators} and of related localization formulas given in terms of Jeffrey-Kirwan (JK) residues on the Coulomb branch \cite{Closset:2015rna, Benini:2015noa, Benini:2016hjo, Closset:2016arn}, which we will briefly review.  In the $O(N)$ case, we will also have to amend those results to account for the non-trivial $\Z_2$ twisted sectors when $g>0$. For instance,  the matching of correlation functions for the $SO(N)/O_+(N')$ duality in \eqref{sd gauge groups} is particularly non-trivial, because of those twisted-sector contributions on the $O_+(N')$ side.

The duality relation \eqref{dual rel} includes some subtle contact terms, which are easily studied by our methods. In particular, the $U(N_c)$ duality involves non-trivial transformations of the Fayet-Iliopoulos (FI) parameters for the global symmetries, which were studied in \cite{Benini:2014mia}.

Finally, let us address the fact that there are two distinct theories with an ``orthogonal gauge group." We must note that, since the group $O(N)$ is in fact disconnected, merely specifying the group does not entirely determine the theory. A convenient way to understand the $O_\pm (N)$ theories is to view them as orbifolds \cite{DHVW1, DHVW2} of an $SO(N)$ gauge theory with $N_f$ chiral multiplets in the vector representation, which has a global $\bZ_2$ symmetry. Depending on $N$ and $N_f$, there may be two distinct orbifolds of a single $SO(N)$ gauge theory \cite{Hori:2011pd}. The states of distinct orbifold theories are obtained by distinct choices of projection in the twisted and untwisted sectors of the theories. These choices, in the genus one partition function, are realized by assigning different weights when summing over $\bZ_2$-twisted partition functions, i.e., partition functions with non-trivial $\bZ_2$ holonomies turned on along the cycles of the torus. Given the choice of weights for the genus-one correlator, the prescription for weighing any $\bZ_2$-twisted partition function is determined, and thus the correlator on any genus-$g$ Riemann surface may be obtained, once the partition functions with non-trivial $\bZ_2$ holonomies are computed. These partition functions, as well as their weighted sums, are computed in sections \ref{sec: SON} and \ref{sec: ON}.

This note is organized as follows. In section \ref{sec: 2}, we summarize some facts about the Coulomb branch of $\CN=(2,2)$ theories, we  discuss the formula \eqref{gen formula correlators} and its relation to the JK residue formula, and we explain how to prove \eqref{dual rel}. In the following sections, we study the dualities \eqref{sd gauge groups} and we prove \eqref{dual rel} in all cases.
The $U(N_c)$ theories are discussed in section \ref{sec: UN};  the $USp(2N_c)$ theories are discussed in section \ref{sec: USpN}; the $SO(N)$ and $O_-(N)$ theories are discussed in sections \ref{sec: SON} and \ref{sec: ON}, respectively.


\section{Coulomb branch correlators on $\Sigma_g$}\label{sec: 2}
Consider a two-dimensional $\CN=(2,2)$ supersymmetric gauge theory, also known as gauged linear sigma model (GLSM), with gauge group $\GG$. Let us denote $\Fg= {\rm Lie}(\GG)$. The theory consists of a $\Fg$-valued vector multiplet
\be
\CV= (a_\mu~,\, \sigma~,\, \t\sigma~,\, \lambda~,\, \t\lambda~,\, D)~,
\ee
 and of chiral multiplets $\Phi_i$ in representations $\FR_i$ of $\Fg$, with standard kinetic terms.
The theory may also have a superpotential $W(\Phi)$ of $R$-charge $2$, which must preserve the vector-like $R$-symmetry $U(1)_R$.
We also have a linear twisted superpotential:
\be\label{CW class}
\CW_0= \sum_I  \tau^I \sigma_I +\sum_F \tau^F m_F~.
\ee
We require that the GLSM preserve the axial-like $R$-symmetry $U(1)_{\rm ax}$, under which $\sigma$ and $m_F$ have charge $2$, at the classical level. This fixes the form of the twisted superpotential \eqref{CW class}. Here we denote by
\be\label{def free center}
\prod_I U(1)_I \subset \GG
\ee
the free part of the center of $\GG$. We define $\sigma_I$ to be the projection of $\sigma$ onto a particular $U(1)_I$ factor, and
\be
\tau^I = {\theta^I \ov 2 \pi} + i \xi^I
\ee
the complexified Fayet-Iliopoulos term for that $U(1)_I$ factor. We also define:
\be\label{def qI}
q_I \equiv e^{2 \pi i \tau_I}~.
\ee
We  couple the background vector multiplets  to the flavor currents in the most general way possible, including the ``flavor'' FI terms $\tau^F$ in \eqref{CW class} (for the abelian part of the flavor group), which lead to contact terms in one-point functions of the conserved current multiplet.~\footnote{It is important to keep track of these contact terms if one is interested in gauging the flavor symmetries. They will also appear in our study of dualities.} The constant value for $\sigma_F$ in the background vector multiplet $\CV_F$, denoted $m_F$, is a familiar ``twisted mass''.

The axial $R$-symmetry can be anomalous in the presence of abelian gauge groups. The $U(1)_{\rm ax}{-}U(1)_I$ anomaly coefficients are:
\be\label{b0I}
b_0^I =\sum_i \Tr_{\FR_i}(t_I)~,
\ee
with $t_I\in i \Fg$ the $U(1)_I$ generator. If $b_0^I=0$ for all $U(1)_I$, the axial $R$-symmetry is preserved quantum-mechanically and the GLSM is expected to flow to a superconformal theory (SCFT) in the infrared. The coefficient \eqref{b0I} is also the one-loop $\beta$-function coefficient for the classically-marginal FI parameter $\tau^I$, with $\mu \d_\mu \tau^I = -{b_0^I\ov 2 \pi i}$.

For any $U(1)_F$ abelian flavor symmetry, we also have the  $U(1)_{\rm ax}{-}U(1)_F$ `t Hooft anomaly coefficients:
\be\label{b0F}
b_0^F = \sum_i Q_i^F {\rm dim}(\FR_i)~,
\ee
with $Q^F_i$ the $U(1)_F$ charge of the chiral multiplet $\Phi_i$.

\subsection{Coulomb branch, twisted superpotential and Bethe vacua}
Consider the classical Coulomb branch of the GLSM, which consists of the constant values:
\be
\sigma= {\rm diag}(\sigma_a)~, \qquad a=1, \cdots, N_c= \rk~,
\ee
for the complex adjoint scalar $\sigma$, breaking the gauge group to its Cartan subgroup $\GH= \prod_a U(1)_a$ modulo the Weyl group $W_\GG$. Let us denote by $\t\fM\cong \C^\rk$ the covering space of the Coulomb branch $\fM= \t\fM\slash W_\GG$.  At a generic point on $\t\fM$ (and for generic values of the twisted masses), the only light fields are the abelian vector multiplets for $\GH$. Integrating out all the massive fields, one obtains the effective twisted superpotential \cite{Witten:1993yc, Witten:1993xi,Nekrasov:2009uh}:
 \be\label{CW eff gen}
 \CW= \CW_0 -{1\ov 2 \pi i}\sum_i \sum_{\rho_i \in \FR_i} (\rho_i(\sigma)+ m_i)\big(\log (\rho_i(\sigma)+ m_i)-1\big)- \half \sum_{\alpha\in \Fg_+} \alpha(\sigma)~,
 \ee
where the sums are over the weights of the representations $\FR_i$ and the positive roots of $\Fg$, respectively.  Here we defined $m_i = Q_i^F m_F$, where the index $F$ runs over the whole flavor group.  Under an axial $R$-symmetry rotation, $\sigma \rightarrow e^{2 i \alpha}\sigma$ and $m_F \rightarrow e^{2 i \alpha}m_F$, the twisted superpotential \eqref{CW eff gen} reproduces the anomalous shifts
\be
\theta^I \rightarrow \theta^I -2 \alpha b_0^I~, \qquad \theta^F \rightarrow \theta^F -2 \alpha b_0^F~,
\ee
of the $\theta$-angles, with the anomaly coefficients given in \eqref{b0I}-\eqref{b0F}.

The so-called Bethe vacua are the solutions to the Bethe equations:
\be\label{Bethe equations}
 \exp{\left(2 \pi i \,{ \d \CW \ov \d \sigma_a}\right)}=1~, \qquad \qquad  w\cdot \sigma\neq \sigma~,\quad \forall w\in W_\GG~,
\ee
for $a=1~, \cdots~, N_c$ running over the Cartan subgroup, modulo the Weyl group action. Here $w\cdot \sigma$, for $w\in W_\GG$, denotes the action of the Weyl group on $\sigma$. The terminology comes from the Bethe/gauge correspondence \cite{Nekrasov:2009uh}.
The second condition in \eqref{Bethe equations} states that an acceptable solution cannot lie on a `Weyl chamber wall' (a locus fixed by the action of  $W_\GG$) in $\t\fM$, where part of the non-abelian gauge symmetry is restored classically.  It is clear that   the approximation that leads to \eqref{CW eff gen} is not valid if $w\cdot \sigma=\sigma$, but it is less clear that there cannot exist additional strongly-coupled ``non-abelian'' vacua at such locations. Following earlier works---in particular the analysis of \cite{Hori:2006dk}---we will {\it assume} this to be true in general: the Bethe vacua give the full set of Coulomb branch vacua.~\footnote{We  have some good circumstantial evidence from localization results for genus zero correlators \cite{Closset:2015rna}. At higher genus, this assumption was made in \cite{Closset:2016arn}, while \cite{Benini:2016hjo} argued for it by using a non-gauge-invariant regulator. See also \cite{Aharony:2016jki}.}

Note that the Bethe equations are always rational equations in the Coulomb branch coordinates $\sigma_a$:
\be\label{Bethe equations explicit gen}
\prod_i \prod_{\rho_i \in\FR_i}(\rho(\sigma)+ m_i^F)^{\rho_i^a} = (-1)^{\sum_{\alpha>0} \alpha^a} \; q_a~, \quad\qquad \qquad w\cdot \sigma\neq \sigma~.
\ee
Here $q_a$ denotes the projection of the FI parameters onto $U(1)_a$.
In theories with only (anti)fundamental flavors, the Bethe equations can be written in terms of a single ``Bethe polynomial''. This is the case for the theories considered in this note.
For future reference, let us also define the Hessian determinant of $\CW$:
\be\label{def H}
H(\sigma)=\det_{ab}\left(-2 \pi i \; \d_{\sigma_a}\d_{\sigma_b} \h W\right) =  \det_{ab}\left(\sum_i \sum_{\rho_i \in \FR_i} {\rho_i^a \rho_i^b \ov \rho(\sigma)+ m_i^F}\right)~,
\ee
which is also a rational function of $\sigma$.

\subsection{Coupling to background fields}
The coupling  to geometric backgrounds of any $\CN=(2,2)$ field theory with a vector $R$-symmetry $U(1)_R$ was studied systematically in \cite{Closset:2014pda, Gomis:2015yaa}, by considering the coupling of the supercurrent to background supegravity \cite{Festuccia:2011ws}. We can preserve two supercharges on any closed oriented Riemann surface $\Sigma_g$ (with $g$ the genus) by the so-called topological $A$-twist \cite{Witten:1988xj}. In addition to the metric, the curved background includes an $R$-symmetry gauge field $A_\mu^{(R)}$ with field strength:
\be
2 i F_{1\b 1} = {1\ov 4} {\rm R}~,
\ee
where ${\rm R}$ is the Ricci curvature.~\footnote{We follow the conventions of  \protect\cite{Closset:2014pda} except that our  definition of ${\rm R}$ differs by an overall sign.}
We therefore have the flux \eqref{flux R} and the $R$-charge quantization condition \eqref{dirac quan R}. In the following, we will consider theories with integer $R$-charges (denoted by $r\in \Z$), which can be coupled to any $\Sigma_g$.

In addition, flavor symmetry currents are naturally coupled to background vector multiplets $\CV_F$, which include background gauge fields $A_\mu^{(F)}$ and background scalars $\sigma_F$. We consider the simplest supersymmetric backgrounds with:
\be\label{V backgd}
{1\ov 2 \pi} \int_{\Sigma_g} d A^{(F)} =  \n_F~,\qquad \qquad\quad
\sigma_F= m_F~,
\ee
with $\n_F$ a GNO-quantized flux (in particular, $\n_F \in \Z$ for a $U(1)_F$ flavor symmetry) and $m_F \in \C$ a constant, the ``twisted mass''.

Note that, on $\Sigma_g$, a mixing of the $R$-symmetry current with a $U(1)_F$ symmetry,
\be\label{mix R to F}
j_\mu^{(R)} \rightarrow j_\mu^{(R)}  + t j_\mu^{(t)}
\ee
is only allowed for $t (g-1)\in \Z$, in order to preserve the Dirac quantization of charges.  This shift is equivalent to a shift of the supersymmetric background flux \eqref{V backgd} by:
\be\label{nf shift}
\n_F \rightarrow \n_F + t (g-1)~,
\ee
with everything else kept constant. The shift \eqref{nf shift} can be understood as a shift of the background vector multiplet:
\be
\CV_F \rightarrow \CV_F + t \CV_R~,
\ee
where $\CV_R$ is an ``R-symmetry vector multiplet'' constructed out of the full supergravity multiplet.~\footnote{See \cite{Closset:2014uda} for a related discussion in higher dimensions.}

The coupling of the GLSM to curved space is conveniently encoded in the ``effective dilaton'' $\Omega= \Omega(\sigma)$, which is the bottom component of a twisted chiral multiplet. The supersymmetric couplings are encoded in the ``improvement Lagrangian'' of  \cite{Closset:2014pda} for $\Omega$, which gives:
\be
\SL_{\Omega}= {i \ov 2} \Omega\;  {\rm R}~,
\ee
when evaluated on the $A$-twist background.
Classically, we may introduce a constant term:
\be
\Omega_0= \tau_R~, \qquad \qquad \tau_R\equiv {\theta_R\ov 2 \pi }+ i \xi_R~,
\ee
which acts as a ``complexified FI parameter'' for $U(1)_R$. In particular, we have:
\be
e^{-\int d^2 x\sqrt{g} \SL_\Omega}= e^{2 \pi i (g-1) \tau_R} \equiv (q_R)^{g-1}~.
\ee
At one-loop on the Coulomb branch, the effective dilaton takes the form \cite{Witten:1993xi, Nekrasov:2014xaa}:
\be\label{Omega eff}
\Omega = \tau_R - {1\ov 2 \pi i}\sum_i \sum_{\rho_i \in \FR_i} (r_i-1)\log (\rho_i(\sigma)+ m_i)
- {1\ov 2 \pi i} \sum_{\alpha \in \Fg}  \log{\alpha(\sigma)}
\ee
with $r_i$ the $R$-charge of the chiral multiplet $\Phi_i$. The last term is the contribution from the $W$-bosons. We therefore have:
\be
e^{2 \pi i \Omega} = q_R \left( \prod_i \prod_{\rho_i\in \FR_i}  (\rho(\sigma) +m_i)^{r_i-1} \prod_{\alpha\in \Fg} \alpha(\sigma)\right)^{-1}~.
\ee

\subsection{Handle-gluing operator and flux operator}
The $A$-twisted theory is a topological field theory \cite{Witten:1988xj}, whose local observables are fully determined by the topological action:
\be\label{S TFT}
S_{\rm TFT}= \int_{\Sigma_g} d^2 x \sqrt{g} \left( -2  {f_{1 \b1}}_a {\d \CW \ov \d \sigma_a} + \t\Lambda^a_{\b1}\Lambda_1^b {\d^2 \CW\ov \d \sigma_a\d\sigma_b} -2  {F_{1 \b1}}^{(F)} {\d \CW\ov \d m_F}   + {i\ov 2} \Omega {\rm R}\right)~,
\ee
which is given in terms of $\CW$ and $\Omega$. Here $f_a= da_a$ and $F^{(F)}=dA^{(F)}$.
We refer to \cite{Closset:2017zgf} for a more thorough discussion.

As in any topological field theory, there exists a local operator $\CH$, the handle-gluing operator, whose insertion corresponds to  ``adding a handle'' to the Riemann surface:
\be
\langle \CO \CH \rangle_g = \langle  \CO \rangle_{g+1}~.
\ee
For $A$-twisted $\CN=(2,2)$ gauge theories, $\CH$ was first computed explicitly in \cite{Nekrasov:2014xaa}---see also \cite{Melnikov:2005tk, Benini:2016hjo, Closset:2016arn}. It is given by:
\be\label{handle gluing op def}
\CH(\sigma)= \exp{\left(2 \pi i \Omega(\sigma)\right)}\; H(\sigma)~,
\ee
where $\Omega$ is the effective dilaton \eqref{Omega eff} and $H$ is the Hessian determinant \eqref{def H}. This latter contribution comes from the gaugino zero-modes in the twisted theory, which couple to $\CW$ as indicated in the second term in \eqref{S TFT}. It is clear from \eqref{S TFT} that $\CH$ corresponds to a local operator insertion one obtains by concentrating the curvature of a single handle at a point, with a $\delta$-function singularity.

Similarly, there exists local operators whose insertion changes the background fluxes for the flavor symmetries. These so-called ``flux operators'' \cite{Closset:2017zgf} are simply given by:
\be\label{def flux op}
\pif_F = \exp{\left(2 \pi i \,{ \d \CW \ov \d m_F}\right)}~,
\ee
in term of the effective twisted superpotential $\CW=\CW(\sigma, m_F)$.

We should also note that the coupling of the GLSM to curved space introduces a ``gravitational'' anomaly for the axial $R$-symmetry $U(1)_{\rm ax}$ \cite{Witten:1993xi, Morrison:1994fr}, with coefficient:
\be\label{def chgrav}
\h c_{\rm grav}  = -{\dim}({\mathfrak g}) - \sum_i (r_i-1) {\dim}({\mathfrak R}_i)~.
\ee
This corresponds to the $U(1)_{\rm ax}{-}U(1)_R$ `t Hooft anomaly:
\be\label{def b0R}
b_0^R = - \h c_{\rm grav}~.
\ee
This anomaly is reproduced by the handle-gluing operator, since
\be
\CH \rightarrow  e^{- 2 i \alpha b_0^R} \CH
\ee
under a $U(1)_{\rm ax}$ rotation, corresponding to an anomalous shift of $\theta_R$. When $b_0^I=0$ and  if the theory flows to a conformal fixed point, $c= 3\,  \h c_{\rm grav}$ is the central charge of the infrared SCFT \cite{Hori:2006dk}.

\subsection{Correlation functions as sums over Bethe vacua}
Let $\CO =\CO(\sigma)$ be a gauge-invariant polynomial in $\sigma$. On the Coulomb branch, this corresponds to a Weyl-invariant polynomial,
\be
\CO(\sigma) \in \C[\sigma_a]^{W_\GG}~.
\ee
The correlation functions of these Coulomb branch operators on $\Sigma_g$ (with background flux $\n_F$) are given explicitly by the formula \cite{Melnikov:2005tk, Nekrasov:2014xaa, Benini:2016hjo, Closset:2016arn}:
\be\label{gen formula correlators bis}
\left\langle \CO(\sigma) \right\rangle_{g; \,\n_F}  = \sum_{\h \sigma\in \CS_{\rm BE}} \CO(\h \sigma) \,\CH(\h\sigma)^{g-1} \, \prod_F \pif_{F}(\h\sigma)^{\n_F}~.
\ee
The sum is over all the distinct solutions $(\sigma_a)= (\h\sigma_a)$ to the Bethe equations \eqref{Bethe equations explicit gen}. Let us note a few simple properties of \eqref{gen formula correlators bis}:
\begin{itemize}
\item It makes the quantum ring relations manifest. The twisted chiral ring relations are the relations $f(\h\sigma)=0$ satisfied by any solution to the Bethe equations, and therefore the insertion of any such relation in the correlation function gives a vanishing result:
\be
\left\langle f(\sigma) \CO(\sigma) \right\rangle_{g; \,\n_F}  =0~.
\ee

\item We easily check that the mixing \eqref{mix R to F} of the $U(1)_R$ symmetry with a flavor symmetry  corresponds to \eqref{nf shift}, as expected. This amounts to a shift of the dilaton effective action by:
\be
\Omega \rightarrow \Omega + t  { \d \CW \ov \d m_F}~.
\ee

\item  Similarly, the mixing of the $R$-symmetry with a gauge symmetry $U(1)_I$ does not change the answer, as expected from gauge invariance. A mixing with the gauge symmetry corresponds to:
\be
\Omega \rightarrow \Omega + t  { \d \CW \ov \d \sigma_I}~,
\ee
but this does not affect $\CH(\h\sigma)$, the handle-gluing operator evaluated on any Bethe vaccum.
\end{itemize}

\subsection{Correlation functions as sums over instantons}\label{subsec: sum over instantons}
It is often interesting to write down the correlation functions in terms of an infinite sum over instanton contributions \cite{Morrison:1994fr}---two-dimensional vortices---in the GLSM:
\be\label{sum instanton gen}
\left\langle \CO(\sigma) \right\rangle_{g; \,\n_F}    = {1\ov |W_\GG|}\sum_{\m \in \Gamma_{\mathbf{G}^\vee}}  q^\m \CZ_{g, \n_F, \m}(\CO)~.
\ee
Here the weight factor $q^\m$ are the FI parameters \eqref{def qI}, the sum is over all GNO-quantized fluxes for $\GG$, and $|W_\GG|$ is the order of the Weyl group. If the free center of $\GG$ \eqref{def free center} is non-trivial, the sum \eqref{sum instanton gen} typically converges for some values of $q_I$, and can be defined more generally by analytic continuation. However, even if $\GG$ does not contain any $U(1)$ factor, we can still make sense of \eqref{sum instanton gen} as a formal sum, which reproduces the prescription \eqref{gen formula correlators bis} after summing over gauge fluxes. Part of the original motivation for this note was to check this claim explicitly, for $USp(2N_c)$ and $SO(k)$ gauge groups.

Recent advances in localization techniques have allowed us to perform that ``microscopic'' computation in  general GLSMs \cite{Closset:2015rna, Benini:2015noa, Benini:2016hjo, Closset:2016arn}---see also \cite{1999math......3178B}. The ``instanton factors''  are given explicitly in terms of Jeffrey-Kirwan (JK) residues on the Coulomb branch covering space $\t\fM$:
\be\label{Zk explicit}
\CZ_{g, \n_F, \m}(\CO) =  \oint_{\rm JK(\eta)} 
 \prod_a \left[ {d\sigma_a \ov 2 \pi i } q_a^{\m_a} \right] \, Z_{g, \n_F, \m}^{\text{1-loop}}(\sigma) \,H(\sigma)^g \, \CO(\sigma)~,
\ee
with $H(\sigma)$ given in \eqref{def H}, and the one-loop determinant:
\bea
&Z_{g, \n_F, \m}^{\text{1-loop}}(\sigma) =\cr
&\qquad\quad  (-1)^{\sum_{\alpha>0}\alpha(\m)} \prod_{\alpha\in \Fg} \alpha(\sigma)^{1-g}\; \prod_i \prod_{\rho_i \in\FR_i} \left(1\ov \rho(\sigma) + m_i\right)^{\rho_i(\m)+\n_i+ (g-1)(r_i - 1)}~.
\eea
Here $m_i$ and $\n_i$ are the  twisted mass and the background flux seen by the chiral multiplet $\Phi_i$, and $r_i$ is its $R$-charge. 
The integration contour in \eqref{Zk explicit} is determined by the Jeffrey-Kirwan prescription with $\eta= \xi_{\rm eff}^{\rm UV}$ for all projective singularities $\sigma_*$ such that $\alpha(\sigma_*)\neq 0$. Here $\xi_{\rm eff}^{\rm UV}\in i\Fh^\ast$ is the effective FI term at infinity on $\t\fM\cong \C^{N_c}$. We refer to  \cite{Closset:2015rna} for more details on the JK residue prescription.

Summing over the fluxes, one can show that \eqref{sum instanton gen} reproduces the Bethe-vacua formula \eqref{gen formula correlators bis}. We will see this in some explicit examples below. We will also see that the result of  \cite{Closset:2015rna, Benini:2015noa} have to be amended in the case of the $O(N)$ gauge group to account for $\Z_2$ twisted sectors, with $\Z_2\cong O(N)/SO(N)$.

\subsection{Matching correlation functions across dualities}\label{duality relations gen}
Consider two theories $\CT$ and $\CT_D$ related by a duality,
\be
\CT \quad\longleftrightarrow \quad\CT_D~.
\ee
 There must be a one-to-one correspondence between Bethe vacua in the dual theories, which means a one-to-one correspondence
between solutions $\h\sigma$ of the Bethe equations in $\CT$ and solutions $\h\sigma^D$ of the Bethe equations in $\CT_D$. By definition, two Coulomb-branch operators $\CO$ and $\CO_D$ are  {\it dual},
\be\label{dual operators in gen}
\CO(\sigma) \in \CT  \quad\longleftrightarrow \quad \CO_D(\sigma^D) \in \CT_D
\ee
if and only if:
\be
\CO(\h\sigma) = \CO_D(\h\sigma^D)~,
\ee
for any pair of dual Bethe vacua $\h\sigma$ and $\h\sigma^D$. To prove the equality \eqref{dual rel} for dual correlators, on any $\Sigma_g$ and with any background flux $\n_F$, we simply need to prove the duality relations:
\be\label{prove dualities gen}
\CH(\h\sigma) = \CH_D(\h\sigma^D)~, \qquad\quad
\pif_F(\h\sigma) = \pif_{F, D}(\h\sigma^D)~,
\ee
for the handle-gluing and flux operators across the duality. For the two-dimensional Seiberg-like dualities that we study in this note, we will see that the equalities \eqref{prove dualities gen} reduce to simple algebraic identities. Three-dimensional dualities have recently been studied with the same methods in \cite{Closset:2016arn, Closset:2017zgf}.


\section{$U(N_c)$ dualities}\label{sec: UN}

Let us consider the $\GG=U(N_c)$ GLSM with $N_f$ chiral multiplets $\Phi_i$ ($i=1, \cdots, N_f)$ and $N_a$ chiral multiplets $\t\Phi^j$ ($j=1, \cdots, N_a$) in the fundamental and antifundamental representations of $U(N_c)$, respectively. We choose the vector-like $R$ symmetry $U(1)_R$ such that:
\be\label{choose R Un}
R[\Phi_i]= r ~, \qquad \qquad R[\t\Phi_j]= \t r~, \qquad \qquad r, \t r\in \Z~.
\ee
Note that we could set $r= \t r$ in flat space by mixing the $R$-symmetry with the gauge symmetry. However, this is not always possible in curved space due to the Dirac quantization condition on the $R$-charge.  We choose the $R$-charges to be integers so that we can consider the theory on a Riemann surface of any genus.~\footnote{Note that we could choose more general $R$-charge $r_i, \t r_j\in \Z$, breaking the flavor group explicitly to its Cartan subgroup. We fix \eqref{choose R Un} for simplicity, and to avoid clutter. The general case can be obtained by mixing the $R$-symmetry with the abelianized flavor symmetry through \eqref{nf shift}.}

This GLSM enjoys a $SU(N_f)\times SU(N_a)\times U(1)_A$ flavor symmetry---see Table \ref{tab:SQCD charges}. We may turn on generic twisted masses $m_i$, $\t m_j$ for the flavor symmetry, with 
\be
\sum_{i=1}^{N_f} m_i = - N_f m_A~, \qquad \qquad \sum_{j=1}^{N_a} \t m_j = N_a m_A~.
\ee
We also consider background flavor fluxes $\n_i, \t\n_j$ on $\Sigma_g$, with $\sum_i \n_i =-N_f\n_A$ and $\sum_j \n_j = N_a \n_A$. 
\begin{table}[t]
\centering
\be\nn
\begin{array}{c|c|ccccc}
    &  U(N_c)& SU(N_f) & SU(N_a)  & U(1)_A &   U(1)_R  \\
\hline
 Q_i   & \bm{{N_c}}  & \bm{\overline{N_f}} &  \bm{1} & 1     &r \\
{\t Q}^j        & \bm{\overline{N_c}}&  \bm{1}&\bm{N_a} & 1      &\t r \\
\end{array}
\ee
\caption{The $U(N_c)$ GLSM gauge, flavor and $R$-charges.}
\label{tab:SQCD charges}
\end{table}

\paragraph{Global anomalies.}
The theory admits a single complexified FI parameter $\tau= {\tau \ov 2 \pi}+ i \xi$ for $U(1) \subset U(N_c)$. It has 
 $\beta$-function coefficient \eqref{b0I} given by:
 \be
b_0 = N_f- N_a~,
\ee
which is also the $U(1)_{\rm ax}$ gauge anomaly.
When $N_f=N_a$, the axial $R$-symmetry survives quantum mechanically and the gauge theory is expected to  have a non-trivial infrared fixed point. 
Let us also note the value of  the `t Hooft anomaly  \eqref{b0F} for $U(1)_A$: 
\be\label{bA UN}
b_0^A = N_c (N_f+ N_a)~.
\ee
The $U(1)_{\rm ax}$ ``gravitational'' anomaly \eqref{def chgrav} is given by:
\be\label{dgrav UN}
\h c_{\rm grav} = \left(N_f (1-r)+ N_a (1-\t r) - N_c\right)N_c~.
\ee

\paragraph{Dual theory.}
This $U(N_c)$ GLSM has an infrared dual description in terms of an $U(N_f-N_c)$ GLSM consisting of $N_a$  fundamental  chiral multiplets $\t q_j$ and $N_f$ antifundamental chiral multiplets $q^i$. The dual theory also contains $N_f N_a$ gauge singlets  ${M^j}_i$ coupled through a superpotential
$W=\t q_j \,{M^j}_i \,q^i$.
The singlets $M^i_j$ are identified with the gauge-invariant mesons $Q^i \t Q_j$ in the original theory.
The flavor and $U(1)_R$ charges are summarized in Table \ref{tab:SQCD dual charges}. The superpotential implies the relation:
\be\label{r rD rel}
r+ \t r+ r_D + \t r_D=2
\ee
between the $R$-charges of the dual theories. The dual theory has a $U(1)_{\rm ax}$ gauge anomaly $b_0^D = - b_0$. We also have the `t Hooft anomalies:
\bea\label{bAD UN}
&b_0^{A,D} &=& -(N_f-N_c)(N_f+N_a) + 2 N_f N_a~, \cr
&\h c_{\rm grav}^D&=& \left(N_f (1-r_D)+ N_a (1-\t r_D) - N_f+N_c\right)(N_f-N_c)\cr
&&&+ N_f N_a (1- r- \t r)~.
\eea
For $N_f=N_a$, the axial $R$-symmetry is an actual symmetry and these anomaly coefficient match:
\be
b_0^{A}= b_0^{A,D}~, \qquad \h c_{\rm grav}= \h c_{\rm grav}^D~, \qquad {\rm if} \quad N_f= N_a~,
\ee
as needed for consistency. For $N_f >N_a$, we find:
\be\label{ref b0 cgrav dual UN}
b_0^{A, D} = b_0^A - N_f b_0~, \qquad\qquad \h c_{\rm grav}^D=   \h c_{\rm grav}- \left(r_D N_f -(r+r_D) N_c\right) b_0~.
\ee
As we will see below, these relations correspond to a non-trivial map of certain contact terms under the duality.

\begin{table}[t]
\centering
\be\nn
\begin{array}{c|c|ccccc}
    &  U(N_f-N_c)& SU(N_f) & SU(N_a)  & U(1)_A &   U(1)_R  \\
\hline
\t q_j   & \bm{{N_f-N_c}}  &  \bm{1}& \bm{\overline{N_a}}  & -1     &\t r_D \\
{ q}^i        & \bm{\overline{N_f-N_c}}& \bm{N_f} & \bm{1}& -1      &r_D \\
{M^j}_i  & \bm{1} &\bm{\overline{N_f}}& \bm{N_a} &  2   & r+\t r \\
\end{array}
\ee
\caption{Charges in the $U(N_f-N_c)$ dual GLSM.}
\label{tab:SQCD dual charges}
\end{table}

\subsection{Twisted chiral ring and duality map}
We are interested in the ring of twisted chiral operators generated by the gauge-invariant polynomials $\Tr(\sigma^p)$, $p=0, \cdots, N_c$, with $\sigma$ the complex scalar in the  $U(N_c)$ vector multiplet.
The structure of the twisted chiral ring can be understood by going onto the Coulomb branch:
\be\label{sigma diag}
\sigma={ \rm diag}\left(\sigma_1~, \cdots~, \sigma_{N_c}\right) = (\sigma_a)~,
\ee
with $a=1, \cdots, N_c$. A convenient basis of twisted chiral operators is given by the elementary symmetric polynomials in $\sigma_a$:
\be\label{def s sym UN}
s_l^{(N_c)}(\sigma) = \sum_{1\leq a_1 < \cdots< a_l \leq N_c} \sigma_{a_1} \sigma_{a_2}\cdots \sigma_{a_l}~, \quad\qquad l=0, \cdots, N_c~,
\ee
Let us define the generating function:
\bea
&Q(z) = \prod_{a=1}^{N_c} (z-\sigma_a) &=& \;\sum_{l=0}^{N_c} (-1)^{l} z^{N_c-l} \; s_l^{(N_c)}(\sigma)\cr
&&=&\; z^{N_c} - z^{N_c-1} \, {\tiny\yng(1)} +  z^{N_c-2}\,  {\tiny\yng(1,1)} -\cdots + (-1)^{N_c} \sigma_1\cdots \sigma_{N_c}~,
\eea
where we identified the symmetric polynomials in $\sigma_a$ with the corresponding Young tableaux.
The twisted chiral ring relations satisfied by the generators \eqref{def s sym UN} are encoded in the effective twisted superpotential $\CW(\sigma)$  of the theory \cite{Hori:2006dk}. We have:
\be
\d_{\sigma_a} \CW= \tau^a -\half (N_c-1) -{1\ov 2\pi i} \left( \sum_{i=1}^{N_f}\log(\sigma_a- m_i)- \sum_{j=1}^{N_a}\log(-\sigma_a-+\t m_j) \right)~,
\ee
modulo an integer.
The Bethe equations are given by:
\be\label{vac eqs}
P(\sigma_a)=0~, \qquad a=1, \cdots N_c~, \qquad\qquad \sigma_a \neq \sigma_b\; \quad{\rm if }\quad a\neq b~,
\ee  
in terms of the polynomial:
\be\label{Pz def Un}
P(z)= \prod_{i}^{N_f} (z - m_i) + (-1)^{N_c} q\, \prod_{j}^{N_a} (-z + \t m_i)~.
\ee
The twisted chiral ring relations can be conveniently written as   \cite{Witten:1993xi, Gaiotto:2013sma, Benini:2014mia}:
\be\label{vac eq 2}
P(z)= C(q) \,Q_D(z) \, Q(z)~,\qquad \qquad C(q)\equiv \begin{cases}
1 &{\rm if}\;\; N_f > N_a~,\\
1+ (-1)^{N_f-N_c} q &{\rm if}\;\; N_f=N_a~,
\end{cases}
\ee
where $Q_D(z)$ is an auxiliary monic polynomial  of degree $N_f-N_c$. The Bethe equations of the dual theory are given by:
\be\label{vac eqs dual}
P(\sigma^D_{\b a})=0~, \qquad {\b a}=1, \cdots N_f- N_c~, \qquad\qquad \sigma^D_{\b a} \neq \sigma^D_{\b b}\; \quad{\rm if }\quad \b a\neq \b b~,
\ee  
in terms of the {\it same} polynomial \eqref{Pz def Un},  where the dual FI parameters are related by:
\be\label{q qD SQCD}
q_D = (-1)^{N_a} q^{-1}~.
\ee
Here we denote by $\sigma^D= (\sigma^D_{\b a})$ the complex scalar of the $U(N_f-N_c)$ vector multiplet on its Coulomb branch. 
Consequently, the polynomial $Q_D(z)$ in \eqref{vac eq 2} should be interpreted as the generating function of the dual twisted chiral ring operators:
\be
Q_D(z) = \prod_{\b a=1}^{N_f-N_c} (z-\sigma_{\b a}^D) = \;\sum_{p=0}^{N_f-N_c} (-1)^{p} z^{N_f-N_c-p} \;s_p^{(N_f-N_c)}(\sigma^D)~.
\ee

The solutions to the Bethe equations \eqref{vac eqs} corresponds to subset of $N_c$ distinct roots of the degree-$N_f$ polynomial $P(z)$. Similarly, the solutions to the dual Bethe equations corresponds to subsets of $N_c-N_f$ distinct roots. Therefore, for any vacua in the original theory, corresponding to a solution  $\{\h \sigma_a\}$, there exists a dual vacua corresponding to the complement $\{\h \sigma_\ba^D\}$ in the set of $N_f$ roots of $P$. Dual operators $\CO(\sigma)$ and $\CO_D(\sigma^D)$ are such that
$\CO(\h \sigma) = \CO_D(\h \sigma^D)$ on any pair of dual vacua.

The relations \eqref{vac eq 2} encode the duality relations between the elementary twisted chiral operators $s_p^{(N_c)}$ and $s_{p'}^{(N_f-N_c)}$ in the dual theories. 
Expanding out \eqref{vac eq 2}, we have $N_f$ equations:
\be
 s_l^{(N_f)}(m) + (-1)^{(N_c+N_a)} q  \, s_{l-N_f+N_a}^{(N_a)}(\t m)=C(q) \sum_{n=0}^l s_{l-n}^{(N_c)}(\sigma) \,s_{n}^{(N_f-N_c)}(\sigma^D)~,
\ee
for $l=1, \cdots, N_f$, where the symmetric polynomials in the twisted masses $m$, $\t m$ are defined like in \eqref{def s sym UN}, with the understanding that $s_l^{(N_a)}=0$ for $l<0$.
Upon solving for the operators  $s_{p'}^{(N_f-N_c)}(\sigma^D)$ in terms of the operators $s_p^{(N_c)}(\sigma)$, we are left with the twisted chiral ring relations of the $U(N_c)$ theory, and vice versa.

\paragraph{Useful identities.} For future reference, let us define:
\bea\label{def F tF}
F(z) \equiv  \prod_{i=1}^{N_f} (z-m_i) = \sum_{l=0}^{N_f}(-1)^l z^{N_f-l} s_l^{(N_f)}(m)~, \cr
\t F(z) \equiv  \prod_{j=1}^{N_a} (z-\t m_j) = \sum_{l=0}^{N_a}(-1)^l z^{N_a-l} s_l^{(N_a)}(\t m)~.
\eea
The polynomial \eqref{Pz def Un} reads:
\be
P(z) = F(z) +(-1)^{N_a+N_c} q\, \t F(z)= C(q) \prod_{\alpha=1}^{N_f}(z-\h z_\alpha)~, 
\ee
where we denote by $\h z_\alpha$ ($\alpha =1, \cdots, N_f$) its $N_f$ roots. We have the useful identities:
\bea\label{useful identity 1}
&\prod_{\alpha=1}^{N_f} (\h z_\alpha-m_i)& =&\;{(-1)^{N_f-N_c} q\ov C(q)} \prod_{j=1}^{N_a} (\t m_j- m_i)~, \cr
&\prod_{\alpha=1}^{N_f}   (\h z_\alpha-\t m_j)&=&\; {(-1)^{N_f} \ov C(q)} \prod_{i=1}^{N_f} (\t m_j- m_i)~.
\eea
Another useful lemma is that, for any partition of the roots $\{\h z_\alpha\}= \{\h \sigma_a \}\cup \{\h \sigma_\ba^D\}$, we have:
\be\label{useful identity 2}
{\prod_{a=1}^{N_c} P'(\h\sigma_a)\ov  \prod_{\substack{a,b=1\\ a\neq b}}^{N_c} (\h\sigma_a-\h\sigma_b)}= (-1)^{N_c(N_f-N_c)} C(q)^{2 N_c-N_f} {\prod_{\ba=1}^{N_f-N_c} P'(\h\sigma_\ba^D)\ov \prod_{\substack{\b a,\b b=1\\\b a\neq\b b}}^{N_f-N_c} (\h \sigma_{\ba}^D-\h\sigma_\bb^D)}
\ee
where $P'(z)= \d_z P(z)$.

\subsection{Equality of correlation functions}
Let us prove the equality of twisted chiral ring correlation functions across the duality, following the strategy of section \ref{duality relations gen}. This proof closely follows similar discussions in \cite{Benini:2013nda, Benini:2014mia, Closset:2016arn}.

\paragraph{Matching the flux operators.} Consider first the flux operators defined by \eqref{def flux op}, for the $SU(N_f)\times SU(N_a)\times U(1)_A$ flavor symmetry.
It is sometimes convenient to consider the decomposition:
\be
m_i = \mu_i - m_A~, \quad \t m_j = \mu_j + m_A~, 
\ee
for the twisted masses, with $m_A$ the $U(1)_A$ twisted mass and $\sum_i \mu_i=0$, $\sum_j= \t\mu_j=0$ for $SU(N_f)\times SU(N_a)$. We similarly decompose the background fluxes as $\n_i = \p_i - \n_A$ and $\t\n_j = \t \p_j + \n_A$.
In the electric theory, the contribution from the flux operators,
\be
\pif_{\rm flux}(\sigma)= \pif_A(\sigma)^{\n_A}\, \prod_{i=1}^{N_f} \pif_i(\sigma)^{\p_i}\, \prod_{j=1}^{N_a} \pif_j(\sigma)^{\t\p_j}~.
\ee
take the simple form:
\be
\pif_{\rm flux}(\sigma)= q^{\n_A}_A (-1)^{N_c N_a \n_A} \prod_{a=1}^{N_c} \left[\prod_{i=1}^{N_f}(\sigma_a-m_i)^{\n_i} \prod_{j=1}^{N_a}(\sigma_a- \t m_j)^{-\t \n_j} \right]~.
\ee
In the dual theory, we find instead:
\bea
&\pif_{{\rm flux}, D}(\sigma)=  q^{\n_A}_{A, D} (-1)^{(N_f-N_c)N_f \n_A} \prod_{\ba=1}^{N_f-N_c}\left[\prod_{i=1}^{N_f}(\sigma_\ba^D-m_i)^{-\n_i} \prod_{j=1}^{N_a}(\sigma_\ba^D- \t m_j)^{\t \n_j} \right]\,\cr &\qquad\qquad\qquad\times\prod_{i=1}^{N_f}\prod_{j=1}^{N_a}(\t m_j-m_i)^{\n_i - \t\n_j}~,
\eea
where the last factor is the contribution from the mesons ${M^i}_j$.
For any pair of dual vacua $\{\h \sigma_a\}$ and $\{\h \sigma_\ba^D\}$, it is easy to see that:
\be
\pif_{\rm flux}(\h\sigma)= \pif_{{\rm flux}, D}(\h\sigma^D)
\ee
follows from the identities \eqref{useful identity 1}, with the non-trivial relation:
\be\label{qA qAD rel}
q_{A, D}=  (-1)^{(N_f-N_c) N_a} q^{-N_f} \, C(q)^{N_f+N_a}\, q_A
\ee
between the $U(1)_A$ flavor contact terms $\tau_A$ and $\tau_{A, D}$ in the dual theories.
Such non-trivial mapping of ``flavor'' FI parameters are related to cluster algebra transformations for two-dimensional supersymmetric quivers \cite{Benini:2014mia}. For $N_f> N_a$, equation \eqref{qA qAD rel} implies the relation:
\be
\xi_{A, D} = \xi_A - N_f \xi
\ee
between flavor FI parameters. This is consistent with the relation \eqref{ref b0 cgrav dual UN} between their one-loop $\beta$-function coefficients, with $b_0^A$ and $b_0^{A, D}$ given in \eqref{bA UN} and \eqref{bAD UN}, respectively.

\paragraph{Matching $\CH$.} Let us consider the handle gluing operator \eqref{handle gluing op def} in the electric theory. The Hessian determinant of $\CW$ is given by:
\be\label{hessian UN}
H(\sigma)=\prod_{a=1}^{N_a} \h H(\sigma_a)~,\qquad \qquad \quad \h H(z)\equiv  \sum_{i=1}^{N_f} {1\ov z-m_i} - \sum_{j=1}^{N_a} {1\ov z - \t m_j}~,
\ee
and the handle-gluing operator reads:
\be
\CH(\sigma) =  q_R \prod_{a=1}^{N_a} \left[ (-1)^{(\t r-1)N_a}\h H(\sigma_a)\ov F(\sigma_a)^{r-1}  \t F(\sigma_a)^{\t r -1} \right]\, \prod_{\substack{a,b=1\\ a\neq b}}^{N_c}  {1\ov \sigma_a-\sigma_b}~,
\ee
in terms if the functions defined in \eqref{def F tF}.
In the dual theory, we have:
\be
\CH_D = h_M \CH_D^{\rm gauge}~,\qquad \qquad h_M = \prod_{i=1}^{N_f}\prod_{j=1}^{N_a} \left(1\ov \t m_j -m_i\right)^{r+ \t r-1}~,
\ee
where $h_M$ is the contribution from the dual mesons, and:
\be
\CH_D^{\rm gauge}(\sigma^D) = q_{R, D} \prod_{\b a=1}^{N_f-N_c} \left[(-1)^{(r_D-1)N_f +1}\h H(\sigma^D_{\b a})\ov F(\sigma^D_{\b a})^{r_D-1}  \t F(\sigma^D_{\b a})^{\t r_D -1} \right]\, \prod_{\substack{\b a,\b b=1\\\b a\neq\b b}}^{N_f-N_c}  {1\ov \sigma_{\ba}^D-\sigma_\bb^D}~.
\ee
is the contribution from all the fields charged under the $U(N_f-N_c)$ gauge group.
Using the fact that 
\be
\d_z P(\h z_\alpha) = \h H(\h z_\alpha) F(\h z_\alpha)~,\qquad \qquad F(\h z_\alpha)= (-1)^{N_c+N_a-1} q \t F(\h z_\alpha)~,
\ee
for any root $\h z_\alpha$, together with the identities \eqref{useful identity 1} and \eqref{useful identity 2}, one can prove that:
\be
\CH(\h \sigma) = \CH_D(\h\sigma^D)~,
\ee
for any dual vacua, with the relation
\be\label{label qR qRD UN}
q_{R, D}= (-1)^{(r+\t r + r_D-1)N_a} q^{r_D N_f - (r+r_D) N_c} \, C(q)^{2(N_c-N_f) + (r+ \t r) N_f} \, q_R 
\ee
between the gravitational contact terms. This is in perfect agreement with the relation  \eqref{ref b0 cgrav dual UN} between the gravitational anomalies. This complete the proof of the equality of dual correlation functions for all the $U(N_c)$ dualities. 

\subsection{Instanton sums and duality relations}
As reviewed in section \ref{subsec: sum over instantons}, the correlation functions can also be written in terms of a sum over instanton contributions. The duality relations imply interesting identities between different JK residues.

\paragraph{Electric theory.}
 The correlation functions of the $U(N_c)$ gauge theory  twisted chiral ring operators $\CO(\sigma)$ on $\Sigma_g$ in the are given by:
\be\label{Zk UNc}
\langle \CO \rangle =q_A^{\n_A}\, q_R^{g-1}  \sum_{\m \in \Z^{N_c}} q^{\sum \m_a}\; \CZ_{g, \m}^{[N_c, N_f, N_a]}(\CO)~,
\ee
The instanton factor is given by the residue integral:
\be\label{Instanton factor UNc}
 \CZ_{g, \m}^{[N_c, N_f, N_a]}(\CO)= {(-1)^{(N_c-1)\sum_a \m_a}\ov N_c!} \oint \prod_{a=1}^{N_c}{d\sigma_a\ov 2 \pi i} \;  \CZ^{\text{1-loop}}_{g, \m}(\sigma)\, H(\sigma)^g \,\CO(\sigma)~,
\ee
with 
\be
\CZ^{\text{1-loop}}_{g, \m} = 
\prod_{a=1}^{N_c}  \left[ \prod_{j=1}^{N_a} (-\sigma_a +\t m_j)^{\m_a-\t\n_j -(g-1)(\t r-1)}\over  \prod_{i=1}^{N_f} (\sigma_a -m_i)^{\m_a- \n_i + (g-1)(r-1)}\right]\,  \prod_{\substack{a,b=1\\ a\neq b}}^{N_c}  (\sigma_a-\sigma_b)^{1-g}~,
\ee
and $H(\sigma)$ given by \eqref{hessian UN}.
The sum in \eqref{Zk UNc} is over  the $U(N_c)$ fluxes $(\m_a)\in \Z^{N_c}$. The contour integral is an iterated residue at all the codimension-$N_c$ singularities of the form:%
~\footnote{Note that the sum over fluxes can be taken as $\m_a \geq M$, with $M$ some integer that depend on the background fluxes $\n_i$ and the $R$-charge $r$.}
\be\label{singularities Un elec}
\sigma_a= m_i^{(a)}~, 
\ee
with $\{m_i^{(a)}\}_{a=1}^{N_c}$ a choice of $N_c$ distinct twisted masses among $\{m_i\}_{i=1}^{N_f}$, and we are assuming that the twisted masses are generic.
The formula \eqref{Zk UNc} follows from \eqref{sum instanton gen}-\eqref{Zk explicit} with $\eta =(1, \cdots, 1)$.~\footnote{Here we assumed that $N_f \geq N_a$. If $N_f>N_a$, this choice of $\eta$ is necessary in order to cancel the contribution from infinity on the Coulomb branch \cite{Closset:2015rna}.} 
The singularities \eqref{singularities Un elec} contribute for $\m_a \geq M$ with $M$ some small-enough integer that depends on the background fluxes $\n_i, \t\n_j$ and on the $R$-charges, and the sum \eqref{Zk UNc} converges for $|q|<1$.

\paragraph{Magnetic theory.} Similarly, the correlation functions of the $U(N_f-N_c)$ dual theory read:
\be\label{Zk UNc dual}
\langle \CO_D \rangle^{\rm dual} =q_{A, D}^{\n_A} \, q_{R, D}^{g-1}  \; Z_M \sum_{\m \in \Z^{N_f-N_c}} q_D^{\sum_{\ba} \m_\ba}\; \t\CZ_{g, \m}^{[N_c, N_f, N_a]}(\CO_D)~,
\ee
where $q_D$ is related to $q$ by \eqref{q qD SQCD}, the $R$-charges are related by \eqref{r rD rel}, and the contact terms are related by  \eqref{qA qAD rel}� and \eqref{label qR qRD UN}.
The factor $Z_M$ in \eqref{Zk UNc dual} is the contribution of the mesons:
\be
 Z_M= \prod_{i=1}^{N_f} \prod_{j=1}^{N_a} \left(1\ov -m_i + \t m_j\right)^{-\n_i + \t\n_j +(g-1)(r+\t r-1)}~,
\ee
and the instanton contribution reads:
\bea\label{JK dual UN}
&\t\CZ_{g, \m}^{[N_c, N_f, N_a]}(\CO_D)=\cr
&\qquad  {(-1)^{(N_f-N_c-1)\sum_{\ba} \m_\ba} \over (N_f-N_c)!} 
   \oint \prod_{\b a=1}^{N_f-N_c}{d\sigma_{\b a}^D\ov 2 \pi i} \; \t\CZ^{\text{1-loop}}_{g, \m}(\sigma^D)\, H_D(\sigma^D)^g \,\CO_D(\sigma^D)~,
\eea
with:
\bea
&\t\CZ^{\text{1-loop}}_{g, \m} = 
\prod_{\b  a=1}^{N_f-N_c}  \left[ \prod_{i=1}^{N_f} (-\sigma^D_\ba + m_i)^{\m_\ba-\n_i -(g-1)(r_D-1)}\over  \prod_{j=1}^{N_a} (\sigma^D_\ba -\t m_j)^{\m_\ba- \t\n_j + (g-1)(\t r_D-1)}\right]\,  \prod_{\substack{\b a,\b b=1\\ \b a\neq \b b}}^{N_f-N_c}  (\sigma^D_\ba-\sigma^D_{\b b})^{1-g}~,\cr
&H_D(\sigma^D)=(-1)^{N_f-N_c} \prod_{a=1}^{N_f-N_c} \h H(\sigma_\ba^D)~,
\eea
with the function $\h H(z)$ defined in \eqref{hessian UN}.
 The contour integral \eqref{JK dual UN}  picks the residues at:
\be
\sigma_\ba^D= m_i^{(\ba)}~, 
\ee
corresponding to a JK residue with $\eta= (-1, \cdots, -1)$ in \eqref{Zk explicit}.

\subsubsection{Integral identities for $N_f >N_a$}
We proved the duality relations:
\be\label{dual rel gen UN}
\langle \CO \rangle=\langle \CO_D \rangle^{\rm dual}~.
\ee
For $N_f >N_a$, a given correlation function receives contribution from a finite number of topological sectors due to the $U(1)_{\rm ax}$ selection rule. Expanding the duality relation \eqref{dual rel gen UN} in $q$, we find the relations:
\bea\label{rel un explicit}
&\sum_{\m_a | \sum_a \m_a= \m_0} \CZ_{g, \m}^{[N_c, N_f, N_a]}(\CO) \cr
&= (-1)^{(N_f-N_c)N_a \n_A + (r+ \t r + r_D-1)N_a (g-1)} \, Z_M   \sum_{\m_\ba | \sum_\ba \m_\ba = \m_0'}\t\CZ_{g, \m}^{[N_c, N_f, N_a]}(\CO_D)~,
\eea
with $\m_0, \m_0' \in \Z$ and
\be
\m_0' = \m_0 + N_f \n_A + (g-1) \left((r+ r_D) N_c - r_D N_f\right)~.
\ee
The sums in \eqref{rel un explicit} are over fluxes that sum to $\m_0$ and $\m_0'$, respectively.
For small values of the parameters, these relations are easily checked on a computer.
We discuss some explicit expressions in Appendix \ref{app: B}.


\section{$USp(2N_c)$ dualities}\label{sec: USpN}
Consider an $\CN=(2,2)$ gauge theory with a gauge group $USp(2N_c)$ and $N_f=2 \kay +1$ flavors. The field content consists of an $USp(2N_c)$ vector multiplet coupled to $N_f$ chiral multiplets $\Phi_i$ ($i=1, \cdots, N_f$) in the fundamental representation, of $R$-charge $r_i \in \Z$. Note that $N_f$ must be odd for the theory to be regular \cite{Hori:2011pd}.
We turn on the twisted masses and fluxes, $m_i$ and $\n_i$, of the $U(1)^{N_f}$ maximal torus of the flavor symmetry group $U(N_f)$. We take the conventions that the chiral multiplet $\Phi_i$ has charge $-1$ in $U(1)_i \subset U(N_f)$.
The proposed dual theory   \cite{Hori:2011pd}  is a  $USp(2N_c^D)$ theory with rank:
\be
N_c^D = \kay - N_c~.
\ee
The dual theory has $N_f$ fundamental chiral fields $\Phi_i^D$ of  $R$-charges:
\be
r_{D,i} = 1-r_i~,
\ee
and inverted flavor charges. It also contains anti-symmetric mesons $M_{ij}$ and a superpotential:
\be
W =  M_{ij} [\Phi_i^D \Phi^D_j]~,
\ee
where the bracket denotes the contraction of the gauge indices with the $USp(2N_c^D)$ invariant two-form.
It follows that the scalar $M_{ij}$ carries $R$-charge $r_i + r_j$.  The fields $M_{ij}$ are identified with the gauge-invariant operators $\t Q_i Q_j$ of the original theory.

The $USp(2N_c)$ theory has a $U(1)_A-U(1)_{\rm ax}$ mixed anomaly, where $U(1)_A$ is the diagonal $U(1)$ in $U(N_f)$, with coefficient:
\be
b_0^A = - 2 N_f N_c~.
\ee
The ``gravitational'' anomaly reads:
\be
\h c_{\rm grav} = - 2N_c \sum_{i=1}^{N_f} (r_i-1)- N_c (2N_c+1) ~,
\ee
with $c=3 \h c_{\rm grav}$ the central charge of the conjectured infrared CFT.
One easily checks that those 't Hooft anomalies are reproduced by the Hori-dual description.

\subsection{Twisted chiral ring and duality map}

The twisted chiral ring of the $USp(2N_c)$ theory can be summarized by a polynomial identity. Just as  with the $U(N)$ duality, it is helpful to consider the two dual theories at once. We introduce the $Q$- and $Q_D$-polynomials, whose coefficients are the gauge-invariant Coulomb branch operators of the $USp(2N_c)$ and $USp(2N_c^D)$ theories, respectively:
\be
Q(z) = \det (z \cdot \mathbf{1} - \sigma) \,, \quad\qquad
Q_D(z) = \det (z \cdot \mathbf{1} - \sigma^D)~.
\ee
The Weyl group of $USp(2N_c)$ is $S_{N_c} \times \Z_2^{N_c}$, which acts on the Cartan coordinates $\sigma_a$ as permutations and sign inversions. Thus, the gauge-invariant twisted chiral operators of $USp(2N_c)$ are given by symmetric polynomials in $\sigma_a^2$:
\be
Q(z) = \det (z \cdot \mathbf{1} - \sigma) = \prod_a (z^2 - \sigma_a^2)~.
\ee
The generators of the classical ring of gauge-invariant twisted chiral operators are given by the coefficients of the $Q$-polynomial. The quantum ring, however, is given by imposing the relations:
\be\label{USp operator map}
P(z) = 2 z Q_D(z) Q(z)
\ee
where
\be\label{def P Usp}
P(z) \equiv \prod_{i=1}^{N_f}(z-m_i) - \prod_{i=1}^{N_f}(-z-m_i) 
= 2z \prod_{\alpha=1}^{\kay}(z^2-\h z_\alpha^2) \,,
\ee
for a set of complex numbers $\h z_{1}, \cdots, \h z_\kay$ defined by the last equation in \eqref{def P Usp}. The quantum relations can be extracted from this equation in an equivalent manner to that explained for the $U(N_c)$ theory. In particular, the operator map can be obtained by expanding the identity \eq{USp operator map} and identifying the coefficients order-by-order in $z$.

By a standard argument, the Bethe vacua of the $USp(2N_c)$ theory, represented by the vacuum expectation value of the Cartan coordinates $\h \sigma_a$, are given by $N_c$-tuples of roots of $P(z)$ that satisfy
\be
\h \sigma_a \neq \pm m_i ~~ \text{for any }a,~i\,,
\quad
\h \sigma_a \neq 0 ~~ \text{for any }a\,,
\quad
\h \sigma_a \neq \pm \h \sigma_b ~~ \text{for }a \neq b\,,
\ee
up to identifications made under the Weyl group. Note that the fact that the root $z=0$ of $P(z)$ must be ignored, due to these constraints, is encoded in the extra factor of $z$ on the right-hand side of equation \eq{USp operator map}. Thus each vacuum can be represented by a $N_c$-tuple
\be
( \h z_{\alpha_1}, \cdots, \h z_{\alpha_{N_c}} )\,, \quad \alpha_1 < \cdots < \alpha_{N_c}\,,
\quad \alpha_a \in [\kay] \,,
\ee
or, more conveniently, by an ascending length-$N_c$ vector of integers:
\be\label{USp alpha}
\alpha=(\alpha_1 ,\cdots,\alpha_{N_c})\,, \quad
\alpha_1 < \cdots < \alpha_{N_c}\,,
\quad \alpha_a \in [\kay] \,.
\ee
Meanwhile, each vacuum in the dual theory can also be represented by a length-$N^D_c$ vector $\alpha^D$. The duality \eq{USp operator map} then implies that the vacuum represented by the vector $\alpha$ in the $USp(N_c)$ theory is mapped to that represented by $\alpha^c$ in the $USp(N^D_c)$ dual theory, where $\alpha^c$ denotes the complement of $\alpha$ within $[\kay]$:
\be
\alpha^D = \alpha^c  = [\kay] \setminus \alpha \,.
\ee

\subsection{$A$-twisted correlation functions}

In this section, we compute the expectation value of operators dual to each other in the mutually dual theories on a genus-$g$ Riemann surface. The expectation value of dual operators match precisely, once we fix a subtle contact term (which corresponds to the relative value of the $U(1)_R$ ``FI parameter'' $\tau_R$ in the dual theories).

To compute the correlator, let us denote the set of vectors $\alpha$ defined in equation \eq{USp alpha} as $\cS(N,\kay)$, i.e.,
\be\label{S}
\cS(N,\kay) = \left\{ (\alpha_1,\cdots,\alpha_N)~:~\alpha_1 < \cdots <\alpha_N\,,~\alpha_a \in [\kay] \right\} \,.
\ee
Then we can express the genus-$g$ partition function of the $USp(2N_c)$ theory as:
\be\label{corr sp gn}
\left\langle \CO(\sigma) \right\rangle_{g; \,\n_F}  = \sum_{\alpha \in \cS(N_c,\kay)} \CO(\h z_\alpha) \,\CH(\h z_\alpha)^{g-1} \, \pif (\h z_\alpha)~.
\ee
We similarly have:
\be\label{corr sp gn D}
\left\langle \CO_D(\sigma^D) \right\rangle_{g; \,\n_F}  =
Z_M\;
\sum_{\alpha^c \in \cS(N_c^D,\kay)} \CO_D (\h z_{\alpha^c}) \,\CH_D(\h z_{\alpha^c})^{g-1} \, \pif_D(\h z_{\alpha^c}) \,,
\ee
in the dual theory,
where we factored out the contribution of the gauge-singlet multiplets $M_{ij}$, which reads:
\be
Z_M =\prod_{1\leq i < j\leq N_f} (-m_i - m_j)^{\n_i + \n_j+ (1-r_i-r_j) (g-1)}~.
\ee
We can easily compute the ratio:
\be\label{USp ratio}
{\CO(\h z_\alpha) \,\CH(\h z_\alpha)^{g-1} \, \pif (\h z_\alpha)} \over
\CO_D (\h z_{\alpha^c}) \,\CH_D(\h z_{\alpha^c})^{g-1} \, \pif_D(\h z_{\alpha^c})~,
\ee
for any $\alpha \in \cS$.  It is useful to note that the Hessian determinant
\be
H(\sigma) = \prod_a \sum_i \left( {1 \over \sigma_a - m_i} - {1 \over \sigma_a + m_i} \right)
\ee
can be simplified using the fact that, for a root $\h z_\alpha$ of $P(z)$, one has:
\be
\sum_i \left( {1 \over \h z_\alpha - m_i} - {1 \over \h z_\alpha + m_i} \right)
= {P'(\h z_\alpha) \over \prod_i (\h z_\alpha -m_i)}
=  {4 \h z_\alpha^2 \prod_{\beta \neq \alpha}(\h z_\alpha^2 - \h z_\beta^2)  \ov  \prod_i (\h z_\alpha - m_i)} \,.
\ee
We arrive at the expressions:
\bea
\pif(\h z_\alpha) &= \prod_i \left[q_A \prod_a (m_i^2 - \h z_{\alpha_a}^2)\right]^{\n_i} \\
\cH(\h z_\alpha) &= q_R
{\prod_{a,\ba} (\h z_{\alpha_a}^2 - \h z_{\alpha^c_\ba}^2)
\ov \prod_{i,a} (m_i^2 - \h z_{\alpha_a}^2)^{r_i-1}
\prod_{i,a} (\h z_{\alpha_a} -m_i)} \,.
\eea
It follows that:
\bea
{\pif(\h z_\alpha) \over \pif_D(\h z_{\alpha^c})}
= \prod_i \left[ q_A q_{A, D}^{-1} \prod_\alpha (m_i^2 - \h z_\alpha^2) \right]^{\n_i}
=(q_A q_{A, D}^{-1})^{\sum_i \n_i}  \prod_{i<j}(-m_i - m_j)^{\n_i+\n_j}
\eea
and
\bea
{\cH(\h z_\alpha) \over \cH_D(\h z_{\alpha^c})}
&= q_R q_{R, D}^{-1} \,
{\prod_{a,\ba}(\h z_{\alpha_a}^2 - \h z_{\alpha^c_\ba}^2)
\prod_{i,\ba} (m_i^2 - \h z_{\alpha^c_\ba}^2)^{-r_i}
\prod_{i,\ba} (\h z_{\alpha^c_\ba} + m_i)
\over
\prod_{a,\ba}(\h z_{\alpha^c_\ba}^2 -\h z_{\alpha_a}^2)
\prod_{i,a} (m_i^2 - \h z_{\alpha_a}^2)^{r_i-1}
\prod_{i,a} (\h z_{\alpha_a} - m_i)} \\
&=q_R q_{R, D}^{-1}\,   e^{i \pi (N_c N^D_c + N_c^D + \nu)}
\prod_{i < j} (-m_i -m_j)^{1-r_i-r_j} \,.
\eea
In appendix \ref{ap: identities}, we show that:
\be
\prod_{i,\alpha}(m_i - \h z_\alpha)
=\prod_{i,\alpha}(m_i + \h z_\alpha)
= e^{i \pi \nu} \prod_{i<j}(m_i + m_j) \,,
\ee 
for an integer $\nu$, uniquely determined by the choice of the masses $(m_1, \cdots, m_{N_f})$.%
\footnote{It is worth noting that, while $\nu$ is independent of the choice of $(\h z_\alpha)$, it shifts by 1 with respect to taking $m_i \rightarrow -m_i$ for all $i$ when $k$ is odd. Note that the polynomial $P(z)$ is invariant under this action.} Finally,  the chiral ring operators map as:
\be
\cO(\h z_\alpha) = \cO_D (\h z_{\alpha^c})~,
\ee
by definition.
The identity of the correlation functions \eqref{corr sp gn}  and \eqref{corr sp gn D} directly follows, with the identifications:
\be
q_{A,D}= q_{A}~, \qquad  q_{R, D}= q_R \, e^{i \pi(N_c N^D_c + N_c^D + \nu)}~,
\ee
between contact terms.

\section{$SO(N)$/$O_+(N)$ dualities}\label{sec: SON}

In this section, we consider theories with $SO(N)$ gauge groups and $N_f$ flavors in the vector representation, and orbifolds thereof. There is a $\bZ_2$ action that acts on the $SO(N)$ group that can be viewed as a global symmetry of the theory, which can be ``gauged," or ``orbifolded" \cite{DHVW1, DHVW2}. This being a discrete symmetry, there are multiple theories that can be obtained by different ways of orbifolding this symmetry. In this section, we consider a particular class of orbifold theories, denoted $O_+$, that are dual to $SO(N)$ theories \cite{Hori:2011pd}.

The matter content of the $SO(N)$ theory is given by $N_f$ chiral multiplets $\Phi_i$ ($i=1, \cdots, N_f$) in the vector representation of $SO(N)$, of $R$-charge $r_i \in \Z$. We consider the twisted masses  $m_i$ and background fluxes $\n_i$ for the $U(N_f)$ flavor symmetry. The chiral multiplet $\Phi_i$ is taken to have charge $-1$ under $U(1)_i \subset U(N_f)$. Finally, note that the $SO(N)$ gauge group admits a $\Z_2$-valued $\theta$ angle, $\theta \in \{0, \pi\}$. In order for the theory to be regular, we need to set $\theta=0$ if $N_f-N$ is odd, and $\theta= \pi$ is $N_f-N$ is even \cite{Hori:2011pd}.

This $SO(N)$ theory is dual to a $O_+(N^D)$ theory with:
\be
N^D = N_f - N+1 \,,
\ee
and $N_f$ chiral fields $\Phi_i^D$ in the vector representation, with $R$-charges
\be
r_{D,i} = 1-r_i
\ee
and inverted flavor charges. The $O_+(N^D)$ theory also contains the symmetric gauge-singlet chiral multiplets $M_{ij}$, which are coupled to the charged chiral multiplets by the superpotential:
\be
W =  (\Phi_i^D)^t M_{ij}  \Phi^D_j~.
\ee
As a simple check, note that the $SO(N)$ theory has 't Hooft anomalies:
\be\label{tHooft SON}
b_0^A = - N N_f~, \qquad\qquad \h c_{\rm grav} =- N \sum_{i=1}^{N_f}(r_i-1)- \half N(N-1)~,
\ee
which are precisely matched by the dual description.

The qualitative description of the duality between $SO(N)$ and $O_+ (N^D_c)$ theories differs depending on the parity of $N$ and $N_f$. We shall describe the duality map and the $A$-twisted correlation functions for each case separately. Before doing so, we first describe the computation of twisted genus-$g$ correlators.

\subsection{Twisted genus-$g$ correlation functions}

In order to compute correlators in an orbifold theory, we must be able to compute correlators with twisted boundary conditions under the orbifold group. In this section, we compute these twisted correlation functions for $\bZ_2$ orbifold theories of $SO$ gauge theories. Since, in the context of this section, the orbifold theories are dual theories of vanilla $SO$ gauge theories, we use notation (superscripts and subscripts on variables and parameters)  convenient for this duality.

We first compute the one-loop determinant $Z_t(\Phi)$ of a chiral field $\Phi$ coupled to the $A$-twisted background and to a background vector multiplet giving rise to an effective twisted mass $m$ and a background flux $\n$, with twisted boundary conditions around certain non-trivial cycles on the Riemann surface, i.e.,
\be
\Phi \rightarrow -\Phi \quad
\text{around cycles $C_1,\cdots,C_k~ (k>0)$ of $\Sigma_g$.}
\ee
We find that:
\be\label{Zt oneloop}
Z_t(\Phi) =  m^{- \n -(r-1)(g-1)}~,
\ee
exactly like for a chiral multiplet in the untwisted sector. 

This can be argued as follows. 
Let us introduce another chiral multiplet $\Phi'$ with the exact same charges, and coupled to the exact same background. We assume, however, that $\Phi'$ is single-valued on $\Sigma_g$. We know the one-loop determinant of $\Phi'$:
\be
Z(\Phi') = m^{-\n-(r-1)(g-1)}~.
\ee
Now we may make the following redefinition of superfields:
\be
\Phi_1 = {1 \ov \sqrt{2}} (\Phi'+\Phi) \,, \quad
\Phi_2 = {1 \ov \sqrt{2}} (\Phi'-\Phi) \,.
\ee
Notice that
\be
\Phi_1 \leftrightarrow \Phi_2 \quad
\text{around cycles $C_1,\cdots,C_k$ of $\Sigma_g$~.}
\ee
Thus the two chiral multiplets can be viewed as a single chiral multiplet living on a double-cover $\Sigma'_{g'}$ of $\Sigma_g$. For this single chiral multiplet, the background flux is given by $2 \n$, while the genus $g'$ is given by $g' = 2g-1$---this is because the integral of the Riemann curvature of $\Sigma'_{g'}$ is double that of $\Sigma_g$, thus $2 - 2g' = 2(2-2g)$.  Meanwhile, the effective twisted mass and the $R$-charge remain the same. Thus, the one-loop determinant of this single chiral multiplet living on $\Sigma'_{g'}$ is given by $ m^{-2 \n -2(r-1)(g-1)}= Z_t(\Phi)Z(\Phi')$.
This implies \eqref{Zt oneloop}.

\subsubsection{Orbifold of $SO(2N_c^D +1)$ theory}

For the $SO(2N_c^D+1)$ gauge theory,  we can fix the gauge such that the orbifolding action, i.e., the generator of the $\bZ_2$ action, acts on the W-bosons $T_{i (2N^D+1)}$, whose matrix elements are given by
\be
(T_{ij})_{kl} = \delta_{ik} \delta_{jl} - \delta_{il} \delta_{kj} \,,
\ee
by an inversion of sign. It also acts on the $(2N_c^D+1)$th component of the fundamental chiral. None of these fields, however, take on vacuum expectation values at the localization locus. Also, as pointed out at the beginning of the section, the one-loop determinants are not affected by twisted boundary conditions. Thus all the twisted sector partition functions agree with the untwisted partition function:
\be\label{SO odd twisted}
\vev{\cO_D }_\text{twisted} = 
\vev{\cO_D }_\text{untwisted}~. 
\ee

\subsubsection{Orbifold of $SO(2N^D)$ theory}

In this case, we can fix the gauge such that the orbifolding action acts on the W-bosons $T_{i, 2N_c}$ by an inversion of sign. It thus acts on the $N_c$-th Cartan element by an inversion. Recall the localization locus, is given by a constant flux and vacuum expectation value for the sigma fields. Since $N$th sigma field and background gauge field must undergo monodromies around cycles of the Riemann surface, it must be that their value is fixed to zero:
\be
\sig^D_{N_c^D} = 0\,, \quad
\m^D_{N_c^D} = 0 \,.
\ee
The generator of $\bZ_2$ also acts on $2N_c^D$-th component of the fundamental chiral, but we know that the one loop determinant of these elements are not modified. The same goes for the W-boson multiplets.

The only remaining problem is to compute the contribution from the light gauginos (or gaugino ``zero modes"). Fortunately, the light gaugino ``mass matrix" does not mix for the $SO$ theory, i.e., $\partial_a \partial_b \hat{W}$ is diagonal. Thus we find that the one-loop determinant for the Cartan elements with indices $a=1,\cdots,N_c-1$ remain the same. We just need to understand what happens for the $N_c$-th Cartan element. Let us denote the one-loop contribution from this element by $Z_{\rm tw}$.

The light ``vector'' gauginos, in an untwisted partition function on a Riemann surface, lie within a multiplet $(a_\mu,\Lambda_z, \bar\Lambda_{\bz})$, with  $a_\mu dx^\mu = \lambda + \bar{\lambda}$, $\Lambda_z = \lambda_z$, $\bar\Lambda_\bz = \bar\lambda_\bz$, for the holomorphic one-form $\lambda$---we follow the notation of \cite{Closset:2015rna, Closset:2016arn}. There are $g$ such one-forms on $\Sigma_g$.

To find the twisted-sector contribution, we again consider the double-cover $\Sigma_{2g-1}$ of the Riemann surface $\Sigma_g$ defined by the twist. Then there are $2g-1$ holomorphic one-forms, as the cover has genus $2g-1$. Now consider the involution $\iota$ that takes a one-form and maps it to a one-form by moving to the alternate cover. Then, by definition, $\iota^2 = \text{id}$. Thus the vector space of holomorphic one-forms decompose into two subspaces, under which $\iota$ acts with eigenvalues $1$ and $-1$, respectively. The one-forms that are invariant under $\iota$ are well-defined on the initial Riemann surface of genus $g$, and are thus holomorphic one forms on $\Sigma_g$.  There are $g$ of them, which we call ``$+$ modes." The number of locally-holomorphic one-forms that satisfy the twisted boundary conditions is given by $(2g-1)-g = (g-1)$. We call them the ``$-$ modes". Note that, at a generic value $\sigma^D$ on the classical Coulomb branch, these modes do not mix, since the mass matrix is invariant under $\iota$. Thus, denoting the one-loop determinant of the $\pm$ modes by $Z_\pm$, we have:
\be
Z_+ Z_-  = Z^\text{one-loop}_{\Sigma_{2g-1}}(\sigma^D) \,.
\ee
Then, by definition,
\be
Z_{\rm tw} = Z_-|_{\sig^D_{N_c^D} = 0} \,.
\ee

Having turned on a generic vev for all of the sigma fields $\sigma^D_\ba$, let us compute the one-loop determinant contribution $Z^\text{one-loop}_{\Sigma_{2g-1}}(\sigma^D) $ of all the light modes on $\Sigma_{2g-1}$. This is given by
\be
Z^\text{one-loop}_{\Sigma_{2g-1}}(\sigma)  = Z(\sig^D_{N_c^D})^{2g-1}:=
(-2\pi i\partial_{N_c^D} \partial_{N_c^D} \hat{W})^{2g-1} \,.
\ee
This one-loop determinant happens to be a function of $\sig^D_{N^D}$ only:~\footnote{Here we take the convention that $\Phi_i$ has charge $1$ under $U(1)_i \subset U(N_f)$. In the conventions of this section, these are the charges of the chiral fields in the dual orbifold theory of the $SO$ theories.}
\be
Z(\sig^D_{N_c^D}) = \sum_{i} \left({1 \ov \sig^D_{N_c^D} + m_{i}} - {1 \ov \sig^D_{N_c^D} - m_{i}}\right) \,.
\ee
Meanwhile, we know the one-loop determinant of the $+$ modes. It is given by
\be
Z_+ = Z(\sig^D_{N_c^D})^{g} \,.
\ee
We thus find:
\be
Z_{\rm tw} = Z_-|_{\sig^D_{N_c^D} = 0} = Z(0)^{g-1} = \left( \sum_{i}{2 \ov m_i} \right)^{g-1} \,.
\ee
We then arrive at the twisted-sector partition function of the $O(2N_c^D)$ gauge  theory:
\bea
&{1 \ov 2^{N^D-1} (N^D-1)!} \sum_{\n^D_\ba}
\oint
\prod_\ba {d \s^D_\ba \ov 2\pi i}
\prod_{\ba \neq \bb} ((\s^D_\ba)^2 -(\s^D_\bb)^2)^{1-g}
\prod_a e^{i \theta \n^D_\ba}\\
&
\quad\cdot \prod_\ba \left[ \sum_{i} \left({1\over \s^D_\ba+m_i}-{1 \over \s^D_\ba-m_i}\right) \right]^{g-1}
\\
&\quad\cdot \prod_i \left(
{(\s_\ba + m_i)^{-\n^D_\ba-\n_i - (g-1) (r_{D,i}-1)}(-\s^D_\ba +m_i)^{\n^D_\ba-\n_i - (g-1) (r_{D,i}-1)} } \right) \\
&\quad\cdot
\left(\sum_{i}{2 \ov m_i} \right)^{g-1}
\prod_{\ba}(-(\s^D_\ba)^4)^{1-g}
\prod_i 
(-m_i)^{2(-\n_i - (g-1) (r_{D,i}-1))}~,
\eea
where the indices $\ba \in [N^D-1]$, {\it not} $[N^D]$. Here $\theta\in \{0, \pi\}$ denotes the $SO(N)$ $\Z_2$-valued $\theta$-angle \cite{Hori:2011pd}. Note that the factor in front of the integral is not equivalent to $|W|^{-1}$, since we have used some of the Weyl symmetry  to fix the Cartan element acted on by the $\bZ_2$ action to be the $N_c^D$-th element.

We can pick the residues of this integrand and arrive at:
\bea\label{SO even twisted}
\vev{\cO_D}_\text{twisted} =
\sum_{\alpha^D \in \cS(N^D-1,\kay)}
\pif_{D,t} (\h z_{\alpha^D}) \cH_{D,t} (\h z_{\alpha^D})^{g-1} \cO_D(\h z_{\alpha^D}) \,,
\eea
for any twisted sector, where we defined:
\bea\label{twisted}
\pif_{D,t} (\h z_{\alpha^D}) &= \prod_i [m_i^2 \prod_\ba (m_i^2 - \h z_{\alpha^D_\ba} ^2)]^{-\n_i} \\
\cH_{D,t} (\h z_{\alpha^D}) &= \prod_\ba (-\h z_{\alpha^D_\ba}^{-2})
\cdot \left( \sum_i {1 \ov 2 m_i} \right) \\
&\quad \cdot
{\prod_i [m_i^2 \prod_\ba (m_i^2 - \h z_{\alpha^D_\ba} ^2) ]^{1-r_{D,i}} \ov 
\prod_{\ba,i}(\h z_{\alpha^D_\ba}+m_i) }
\cdot
{\prod_\ba P'(\h z_{\alpha^D_\ba}) \ov
\prod_\ba \h z^2_{\alpha^D_\ba} \prod_{\ba \neq \bb} (\h z_{\alpha^D_\ba}^2-\h z_{\alpha^D_\bb}^2)} \,.
\eea
The polynomial $P(z)$ will be defined in section \ref{subsec: ON dualities and tcr}.
Here, $\alpha^D$ is a vector of length $N_c^D-1$ with
\be\label{SOtwisted alpha}
\alpha^D=(\alpha^D_1 ,\cdots,\alpha^D_{N_c^D-1})\,, \quad
\alpha^D_1 < \cdots < \alpha^D_{N_c^D-1}\,,
\quad \alpha^D_\ba \in [\kay] \,,
\ee
i.e., elements of $\cS(N_c^D-1,\kay)$. $\kay$ is defined so that the number of non-zero roots of $P(z)$ is $2 \kay$. $\h z_\alpha$ are the non-zero roots of $P(z)$:
\be
P (z) =
\begin{cases}
2 z \prod_{\alpha=1}^\kay (z^2 -\h z_\alpha^2) & \text{when $N_f = 2\kay+1$} \\
2 \prod_{\alpha=1}^\kay (z^2 -\h z_\alpha^2)& \text{when $N_f = 2\kay$} \,.
\end{cases}
\ee
When $N_f = 2 \kay+1$, it is useful to note that a pole $(\sigma^D_\ba)  =(\pm \h z_{\alpha^D_1}, \cdots, \pm\h z_{\alpha^D_{N_c^D-1}})$ of the integrand corresponds to a vacuum represented by the set of roots $(\h z_{\alpha^D_1}, \cdots,\h z_{\alpha^D_{N_c^D-1}},0)$ in the $O(2N_c^D)$ theory, while there is no such interpretation when $N_f$ is even.

\subsection{Summing over the twisted sectors}

In order to obtain an $A$-twisted correlator on $\Sigma_g$ of an orbifold theory, we must sum over the correlators computed in the twisted sectors. In order to label the twisted sectors, let us denote the $g$ $A$-cycles and $g$ $B$-cycles of the Riemann surface by $A_I$ and $B_I$ such that
\be
A_I \cdot A_J = B_I \cdot B_J = 0 \,, \quad
A_I \cdot B_J = \delta_{IJ} \,.
\ee
Restricting the orbifold group to be $\bZ_2$, a twisted sector is labeled by the cycles the $\bZ_2$ twist is applied on:
\be
\{ A_{I_1}, A_{I_2}, \cdots ,B_{J_1}, B_{J_2}, \cdots \} \,.
\ee
A consistent prescription of adding the partition functions to compute a vacuum expectation value is to weigh each twisted partition function by $v^{g-N(T)} w^{N(T)}$, where
\be
N(T) = (\text{number of indices $I$ such that either $A_I \in T$ or $B_I \in T$}) \,,
\ee
for some constant $w$. This prescription lead to invariance under $B_I \rightarrow B_I + A_I$, $B_I \rightarrow A_I$ and $(A_I,B_I) \leftrightarrow (A_J,B_J)$ for $I \neq J$. The values of $v$ and $w$ depend on the choice of the orbifold projection we take.

A simple way of understanding these weights is by considering the genus-one partition function. There, the choice of orbifold projection leads to a prescription of $v$ and $w$ for each partition function with holonomies of the orbifold action $\Gamma$, as discussed in \cite{Hori:2011pd}. For example, in the case that the orbifold projection is such that the untwisted sector is projected down to the $\Gamma$, $v$ and $w$ are taken to be 1 \cite{DHVW1, DHVW2}.~\footnote{Such orbifolds, and their genus-one partition functions have been reviewed in \cite{2DReview}.}  Once these weights are determined, they can be used to sum over higher-genus partition functions. A heuristic way of understanding this prescription is to recall that the genus-$g$ correlators can be viewed as correlators on the sphere with $g$ insertions of handle operators. Each handle operator is realized by introducing the handle, and summing over all possible holonomies on each cycle of the handle with a prescribed weight. From this point of view, it is trivial that these prescribed weights should be identified with the weights with which the genus-one partition functions are summed.

Now in the previous subsection, we have shown that the vacuum expectation value of an operator only depends on whether there exists a cycle with a non-trivial $\bZ_2$ twist or not. That is, for any nonempty $T$,
\be
\vev{\cO_D}_{T} = \vev{\cO_D}_{\text{twisted}} \,.
\ee
Thus the $A$-twisted expectation value is given by
\bea
\vev{\cO_D}
&={1 \ov |\bZ_2|} \sum_T v^{g-N(T)} w^{N(T)} \vev{\cO_D}_T \\
&={1\over 2} v^g \vev{\cO_D}_\text{untwisted} + {1\over 2} \left( \sum_{N(T)=1}^g
\begin{pmatrix}
g \\ N(T)
\end{pmatrix}
3^{N(T)} v^{g-N(T)} w^{N(T)} \right) \vev{\cO^D}_\text{twisted} \\
&= {1\over 2} v^g \vev{\cO_D}_\text{untwisted}
+{1\over 2} \Big( (v+3w)^{g} - v^g\Big) \vev{\cO_D}_\text{twisted} \,.
\eea

\subsection{Twisted chiral ring and dualities}\label{subsec: ON dualities and tcr}

The elements of the twisted chiral ring of the $SO(N)$ theory can be represented by the Weyl-invariant polynomials of the sigma-fields $\sigma_a$. When $N$ is odd, these are just symmetric polynomials of $\sigma_a^2$, the generators thus being the elementary symmetric polynomials of $\sigma_a^2$, whose generating function is given by
\be
Q(z) = \det(z \cdot \mathbf{1} - \sigma) \,.
\ee
Meanwhile, when $N$ is even, there is an additional generator of the twisted chiral ring, being the Pfaffian of $\sigma$:
\be
\text{Pf} (\sigma) = \prod_a \sigma_a \,.
\ee
This is because, for $SO(2N)$, the Weyl group consists of permutations of $\sigma_a$ and sign inversions $\epsilon_a$ on $\sigma_a$ which satisfies $\prod_a \epsilon_a = 1$. In the orbifold theory, whichever orbifold one chooses to take, the gauge invariant local operators are given by symmetric polynomials of $(\sigma^D_\ba)^2$. The generating function for the elementary symmetric polynomials, again, is given by
\be
Q_D (z) = \det(z \cdot \mathbf{1} - \sigma^D) \,. 
\ee
Depending on the orbifold projection, however, there may be a twist field $\tau$ in the twisted chiral ring.

The quantum twisted chiral ring of the $SO(N)$ theory is then summarized by
\be\label{SO operator map}
zP(z) = 2 Q_D(z) Q(z) \,,
\ee
where the polynomial $P(z)$ is defined to be
\be
P(z) = \prod_{i=1}^{N_f} (z - m_i) + \prod_{i=1}^{N_f} (z+m_i) \,.
\ee
These twisted chiral ring relations directly follow from the twisted superpotential of the $SO(N)$ theory, with the $\Z_2$ $\theta$-angle taken to be trivial if $N-N_f$ is odd, and with $\theta=\pi$ if $N-N_f$ is even \cite{Hori:2011pd}.

When $N$ is even, there is an additional (trivial) relation one needs to take in to account:
\be
\text{Pf}(\sigma)^2 = \prod_a \sigma_a^2 \,.
\ee
Thus, when $N$ is even, in the dual $O_+(N^D)$ theory, there is a twist operator $\tau$ corresponding to the Pfaffian operator in the twisted chiral ring. Note that the dual operator of $\prod_a \sigma_a^2$ is a symmetric polynomial of $(\sigma_\ba^D)^2$ of degree $N_c$. We denote this symmetric polynomial by $(\prod_a \sigma_a^2)^D$. Then the twist operator satisfies the relation:
\be
\tau^2 = \left( \prod_a \sigma_a^2 \right)^D \,.
\ee
The description of the twisted chiral vacua, and the evaluation of the correlation functions, vary qualitatively depending on the parity of $N$ and $N^D$. We now proceed to describe these features in each case.

\subsubsection{$SO(2N_c) \leftrightarrow O_+(2N_c^D+1)$, $N_f = 2 \kay$, $N_c^D = \kay -N_c$}

\noindent\textbf{Map of vacua :}
The number of flavors being even, $P(z)$ can be written as
\be
P(z) = 2\prod_{\alpha=1}^\kay (z^2 - \h z_\alpha^2)
\ee
The Coulomb branch vacua of the $SO(2N_c)$ theory, represented by the vacuum expectation value of the Cartan coordinates $\h \sigma_a$, are given by $N_c$-tuples of roots of $P(z)$ that satisfy certain constraints.
\be
\h \sigma_a \neq \pm m_i ~~ \text{for any }a,~i\,,
\quad
\h \sigma_a \neq \pm \h \sigma_b ~~ \text{for }a \neq b\,,
\ee
up to identifications made under the Weyl group. There are then two sets of vacua:
\bea\label{Nc tuples doublets}
&( \h z_{\alpha_1}, \cdots, \h z_{\alpha_{N_c}} )\,, \quad& \alpha_1 < \cdots < \alpha_{N_c}\,,
\qquad& \alpha_a \in [\kay] \,,\cr
&( \h z_{\alpha_1}, \cdots, \h z_{\alpha_{N_c-1}},-\h z_{\alpha_{N_c}} )\,, \quad &\alpha_1 < \cdots < \alpha_{N_c}\,,
\qquad& \alpha_a \in [\kay] \,.
\eea
That is, for each $\alpha\in \cS(N_c,\kay)$, an ascending length-$N_c$ vector:
\be\label{SO alpha}
\alpha=(\alpha_1 ,\cdots,\alpha_{N_c})\,, \quad
\alpha_1 < \cdots < \alpha_{N_c}\,,
\quad \alpha_a \in [\kay] \,,
\ee 
there are two associated vacua.

Likewise, in the dual theory, two vacua can be associated to an ascending length-$N_c^D = (\kay-N_d)$ vector $\alpha^D$. In this case, there is only one representative $N_c^D$-tuple of roots:
\be
( \h z_{\alpha^D_1}, \cdots, \h z_{\alpha^D_{N^D}} )\,, \quad \alpha^D_1 < \cdots < \alpha^D_{N^D}\,,
\quad \alpha^D_\ba \in [\kay] \,,
\ee
corresponding to $\alpha^D$. However, viewed as the vacuum expectation value of $\sigma^D$, this is a fixed point of the $\Z_2$ orbifold action. The $O_+ (2N_c^D +1)$ theory is defined so that both the twisted and untwisted states corresponding to this vacuum expectation value are included in the twisted-chiral spectrum. The two vacua represented by the vector $\alpha$ in the $SO(2N_c)$ theory are mapped to those corresponding to $\alpha^D= \alpha^c$ in the $O_+(2N_c^D+1)$ dual theory, where $\alpha^c$ denotes the complement of $\alpha$ within $[\kay]$.
\medskip

\noindent\textbf{$A$-twisted correlation functions :} The $A$-twisted correlation function of the $SO(2N_c)$ theory is given by:
\be\label{orig SO 2N 2k}
\vev{\cO_0 + \text{Pf} (\sigma) \cdot \cO_1 }_{g;\n_F} = 
2 \sum_{\alpha \in \cS(N_c,\kay)}  \cO_0 (\h z_\alpha) \cH(\h z_\alpha)^{g-1} \pif (\h z_\alpha)
\ee
where we have decomposed an arbitrary operator $\cO$ into
\be
\cO = \cO_0 + \text{Pf} (\sigma) \cdot \cO_1 \,,
\ee
where $\cO_0$ and $\cO_1$ are polynomials of $\sigma_a^2$. We have:
\bea\label{SO 2N}
\pif(\h \sigma_a) &=q_A^{\sum_i \n_i} \prod_{i,a} (m_i^2 -\h \sigma_a^2)^{\n_i} \\
\cH (\h \sigma_a) &=q_R
{\prod_{i,a} (m_i^2 -\h \sigma_a^2)^{1-r_i}  \over \prod_{i,a} (\h \sigma_a - m_i)}
\cdot {\prod_a P'(\h \sigma_a) \over \prod_{a \neq b} (\h \sigma_a^2 - \h \sigma_b^2)} \,,
\eea
when $\h\sigma_a$ are roots of $P(z)$.

The sum over $\cS(N_c,\kay)$ and the projection of the operator to $\cO_0$ should be commented on. Recall that there are two vacua of the $SO(2N_c)$ theory corresponding to each element of $\cS(N_c,\kay)$. By picking up the poles of the summed integrand as before, we find that the vacuum expectation value of an arbitrary operator $\cO$ can be written as:
\bea\label{pre orig SO 2N 2k}
&\vev{\cO}_{g;\n_F} =
\sum_{\alpha \in \cS(N_c,\kay)}  \cO (\h z_{\alpha_1},\cdots,\h z_{\alpha_N})
\cH(\h z_{\alpha_1},\cdots,\h z_{\alpha_N})^{g-1} \pif (\h z_{\alpha_1},\cdots,\h z_{\alpha_N}) \\
&\qquad+ \sum_{\alpha \in \cS(N_c,\kay)}  \cO_0 (\h z_{\alpha_1},\cdots,-\h z_{\alpha_N})
\cH(\h z_{\alpha_1},\cdots,-\h z_{\alpha_N})^{g-1} \pif (\h z_{\alpha_1},\cdots,-\h z_{\alpha_N}) \,.
\eea
Now using the identities of appendix \ref{ap: identities}, we find that
\be
\pif(\h z_{\alpha_1},\cdots,\h z_{\alpha_N})=\pif(\h z_{\alpha_1},\cdots,-\h z_{\alpha_N})\,, \quad
\cH(\h z_{\alpha_1},\cdots,\h z_{\alpha_N})=\cH(\h z_{\alpha_1},\cdots,-\h z_{\alpha_N})\,,
\ee
while, by definition,
\be
\cO_0 \rightarrow \cO_0 \,, \qquad
\cO_1 \rightarrow \cO_1 \,, \qquad
\text{Pf}(\h \sigma_a) \rightarrow -\text{Pf} (\h \sigma_a) \,.
\ee
under $\h z_{\alpha_N} \rightarrow -\h z_{\alpha_N}$. Thus the expectation value \eq{pre orig SO 2N 2k} is given by equation \eq{orig SO 2N 2k}.

Meanwhile, in the dual theory, we find that
\be
\vev{\cO_{0,D} + \tau \cdot \cO_{1,D}}_{g:\n_F} = \vev{\cO_{0,D}}_{g:\n_F} \,,
\ee
since a single twist operator introduces a single branch cut, thus its expectation value must vanish. Let us also note that, for an orbifold of the $SO(2N_c^D +1)$ theory, the twisted sector expectation value coincides with the untwisted expectation value, leading to:
\be
\vev{\cO_{0,D}}_{g:\n_F} 
= {1 \ov 2} (v+3w)^g \vev{\cO_{0,D}}_{g:\n_F,\text{untwisted}}\,,
\ee
for parameters $v$ and $w$, which depend on the orbifold projection. We finally arrive at
\bea
\vev{\cO_{0,D}+\tau \cdot \cO_{1,D}}_{g:\n_F} &= {(v+3w)^g \ov 2}
Z_M 
\sum_{\alpha^D \in \cS(N_c^D,\kay)}
\cO_{0,D} (\h z_{\alpha^D}) \cH_D(\h z_{\alpha^D})^{g-1} \pif_D (\h z_{\alpha^D})~,
\eea
with
\bea\label{O 2ND+1}
\pif_D(\h \sigma^D_\ba) &=q_{A,D}^{\sum_i \n_i} \prod_i \left( m_i \prod_{\ba} (m_i^2 -(\h \sigma^D_\ba)^2) \right)^{-\n_i} \\
\cH_D (\h \sigma^D_\ba) &=q_{R,D}
{\prod_i \left( m_i \prod_{\ba} (m_i^2 -(\h \sigma^D_\ba)^2) \right)^{1-r_{D,i}} \over 
\prod_{i,\ba} (\h \sigma^D_\ba + m_i)}
\cdot {\prod_\ba P'(\h \sigma^D_\ba) \over \prod_\ba \h \sigma_\ba^2
\prod_{\ba \neq \bb} ((\h \sigma^D_\ba)^2 - (\h \sigma^D_\bb)^2)}
\eea
the contribution from the dual gauge theory, where $\h \sigma_\ba$ are roots of $P(z)$, and 
\be\label{ZM SO}
Z_M = \prod_{i \leq j} (-m_i -m_j)^{\n_i + \n_j + (1-r_i -r_j) (g-1)} 
\ee
the contribution from the meson singlets.
Using the identities listed in appendix \ref{ap: identities}, we then find that
\bea
{\pif (\h z_\alpha) \over \pif_D (\h z_{\alpha^c})}
&= (q_A q_{A,D}^{-1})^{\sum_i \n_i} \,2^{-2 \sum_i \n_i} e^{i \pi \sum_i \n_i} \prod_{i \leq j}(-m_i -m_j)^{\n_i + \n_j} \\
{\cH (\h z_\alpha) \over \cH_D (\h z_{\alpha^c})}
&= q_R q_{R,D}^{-1}\,2^{-4N_c^D+2 \sum_i r_i} e^{i \pi(N_cN_c^D+\nu+ \sum_i r_i)} \prod_{i \leq j}(-m_i -m_j)^{1-r_i -r_j} \,.
\eea
Also, by the operator map \eq{SO operator map},
\be
\cO_0 (\h z_\alpha) = \cO_{0,D} (\h z_{\alpha^c}) \,,
\ee
$\cO_0$ being symmetric polynomials of the square $\sigma_a^2$ of the Cartan coordinates of $\sigma$. We thus find that
\bea
&\vev{\cO_0 + \text{Pf}(\sigma) \cO_1}_{g;\n_F}
=  4 (v+3 w)^{-g}  (q_A q_{A,D}^{-1})^{\sum_i \n_i}(q_R q_{R, D}^{-1})^{g-1}
\\
&\cdot e^{-(2 \ln 2 + i \pi) \sum_i \n_i } e^{\left[ -(4 \ln 2) N_c^D + i \pi (N_cN_c^D + \nu)+ (2 \ln 2 +i \pi) \sum_i r_i \right](g-1)}   \vev{\cO_{0,D} + \tau \cO_{1,D}}_{g;\n_F} \,.
\eea
With the prescription $v=w=1$---that is, adding up all the twisted sectors with weight 1---we arrive at
\be
\vev{\cO_0 + \text{Pf}(\sigma) \cO_1}_{g;\n_F}
= \vev{\cO_{0,D} + \tau \cO_{1,D}}_{g;\n_F} \,,
\ee
with the identifications:
\be
q_{A,D}= e^{-(2 \ln 2 + i \pi)}\, q_A~, \quad q_{R,D}=e^{\left[ -(2 \ln 2) (2N_c^D+1) + i \pi (N_cN_c^D + \nu)+ (2 \ln 2 +i \pi) \sum_i r_i \right]}\, q_R~,
\ee
amongst the contact terms.

\subsubsection{$SO(2N_c) \leftrightarrow O_+(2N_c^D)$, $N_f = 2 \kay+1$, $N_c^D = \kay -N_c+1$}

\noindent\textbf{Map of vacua :}
$P(z)$ is given by
\be
P(z) = 2z \prod_{\alpha=1}^\kay (z^2 - \h z_\alpha^2)~.
\ee
There are three types of vacua in the $SO(2N_c)$ theory. First, we have the two sets of vacua that can be represented by $N_c$-tuples like in \eqref{Nc tuples doublets}. That is:
\bea
&( \h z_{\alpha_1}, \cdots, \h z_{\alpha_{N_c}} )\,, \quad& \alpha_1 < \cdots < \alpha_{N_c}\,,
\qquad& \alpha_a \in [\kay] \,,\cr
&( \h z_{\alpha_1}, \cdots, \h z_{\alpha_{N_c-1}},-\h z_{\alpha_{N_c}} )\,, \quad &\alpha_1 < \cdots < \alpha_{N_c}\,,
\qquad& \alpha_a \in [\kay] \,.
\eea
This gives two vacua associated to each element of $\cS(N_c,\kay)$. In addition, there are vacua represented by a $(N_c-1)$-tuples:
\be
( \h z_{\alpha_1}, \cdots, \h z_{\alpha_{N_c-1}},0 )\,, \qquad \alpha_1 < \cdots < \alpha_{N_c-1}\,,
\qquad \alpha_a \in [\kay] \,.
\ee
These vacua are in one-to-one correspondence with elements $\alpha \in \cS(N_c-1,\kay)$.

In the dual theory, two vacua can be associated to an ascending length-$(N_c^D-1) = (\kay-N_c)$ vector $\alpha^D \in \cS(N_c^D-1,\kay)$. The representative tuples of roots are given by:
\be
( \h z_{\alpha^D_1}, \cdots, \h z_{\alpha^D_{N_c^D-1}},0 )\,, \quad \alpha^D_1 < \cdots < \alpha^D_{N_c^D-1}\,,
\quad \alpha^D_\ba \in [\kay] \,,
\ee
corresponding to $\alpha^D$. This vacuum expectation value of $\sigma^D$, being a fixed point of the orbifold action, has two vacua associated to it, according to the definition of $O_+(2N_c^D)$. Meanwhile, there is a single vacuum for each tuple of roots
\be
( \h z_{\alpha^D_1}, \cdots, \h z_{\alpha^D_{N_c^D-1}},0 )\,, \quad \alpha^D_1 < \cdots < \alpha^D_{N_c^D-1}\,,
\quad \alpha^D_\ba \in [\kay] \,,
\ee
such a tuple not being a fixed point of the orbifold action. There vacua are in one-to-one correspondence with elements $\alpha^D\in \cS(N^D,\kay)$.

The duality map of the twisted ground states, as before, is given by taking complement of a vector $\alpha$ representing vacua of the $SO(2N_c)$ theory with respect to $[\kay]$. One finds that:
\bea\label{sumup dual rel 532}
&\alpha \in \cS(N_c,\kay)~ &\Leftrightarrow ~&\quad\alpha^c \in \cS(N_c^D-1,\kay)\,, \\
&\alpha \in \cS(N_c-1,\kay)~ &\Leftrightarrow ~&\quad\alpha^c \in \cS(N_c^D,\kay) \,.
\eea
There are two vacua per vector in the first line, while there is one vacuum per vector in the second line.

\medskip

\noindent\textbf{$A$-twisted correlation functions :} The $A$-twisted correlation function of the $SO(2N_c)$ theory is given by
\bea\label{orig SO 2N 2k+1}
\vev{\cO_0 + \text{Pf} (\sigma) \cdot \cO_1 }_{g;\n_F} = 
&2 \sum_{\alpha \in \cS(N_c,\kay)}  \cO_0 (\h z_\alpha) \cH(\h z_\alpha)^{g-1} \pif (\h z_\alpha) \\
&+ \sum_{\alpha \in \cS(N_c-1,\kay)}  \cO_0 (\h z_\alpha,0) \cH(\h z_\alpha,0)^{g-1} \pif (\h z_\alpha,0)
\eea
where, as before, an arbitrary operator $\cO$ has been decomposed in to
\be
\cO = \cO_0 + \text{Pf} (\sigma) \cdot \cO_1 \,,
\ee
with $\cO_0$ and $\cO_1$ being polynomials of $\sigma_a^2$. The notation is such that
\be
\pif(\h z_\alpha) = \pi(\h z_{\alpha_1} ,\cdots,\h z_{\alpha_{N_c}})\,, \qquad
\pif(\h z_\alpha,0) = \pi(\h z_{\alpha_1} ,\cdots,\h z_{\alpha_{N_c-1}},0)\,,
\ee
and similarly for $\cH$ and $\cO$. The operators $\pif$ and $\cH$ are given by equation \eq{SO 2N}. The factor of 2 in the first term of equation \eq{orig SO 2N 2k+1} and the projection to $\cO_0$ for the vacua represented by $\alpha \in \cS(N_c,\kay)$ has been commented on previously. Note that for $\alpha \in \cS(N_c-1,\kay)$, we have $\text{Pf}(\h z_\alpha,0) =0$.

In the dual theory, as before, 
\be
\vev{\cO_{0,D} + \tau \cdot \cO_{1,D}}_{g;\n_F} = \vev{\cO_{0,D}}_{g;\n_F} \,.
\ee
The untwisted partition function is then given by
\bea
{\vev{\cO_{0,D}}_{g;\n_F,\text{untwisted}} \over Z_M}=
&\sum_{\alpha^D \in \cS(N_c^D-1,\kay)}
\cO_{0,D} (\h z_{\alpha^D},0)
\cH_{D} (\h z_{\alpha^D},0)^{g-1}
\pif_{D} (\h z_{\alpha^D},0) \\
&+2 \sum_{\alpha^D \in \cS(N_c^D,\kay)}
\cO_{0,D} (\h z_{\alpha^D})
\cH_{D} (\h z_{\alpha^D})^{g-1}
\pif_{D} (\h z_{\alpha^D})
\eea
for
\bea\label{O 2ND}
\pif_D(\h \sigma^D_\ba) &= q_{A,D}^{\sum_i \n_i}\prod_{i,\ba} (m_i^2 -(\h \sigma^D_\ba)^2)^{-\n_i} \\
\cH_D (\h \sigma^D_\ba) &=q_{R,D}
{\prod_{i,\ba} (m_i^2 -(\h \sigma^D_\ba)^2)^{1-r_{D,i}}  \over \prod_{i,\ba} (\h \sigma_\ba + m_i)}
\cdot {\prod_\ba P'(\h \sigma^D_\ba) \over \prod_{\ba \neq \bb} (\h (\sigma^D_\ba)^2 - (\h \sigma^D_\bb)^2)} \,.
\eea
Note that we have factored out the meson determinant $Z_M$.  We also find that the vacuum expectation value in the twisted sectors:
\bea
{\vev{\cO_{0,D}}_{g;\n_F,\text{twisted}} \over Z_M}=
&\sum_{\alpha^D \in \cS(N_c^D-1,\kay)}
\cO_{0,D} (\h z_{\alpha^D})
\cH_{D,t} (\h z_{\alpha^D})^{g-1}
\pif_{D,t} (\h z_{\alpha^D})
\eea
for $\pif_{D,t}$ and $\cH_{D,t}$ defined in equation \eq{twisted}. Quite non-trivially, we find that:
\be
\cH_{D,t} (\h z_{\alpha^D}) = \cH_D (\h z_{\alpha^D},0) \,, \quad
\pif_{D,t} (\h z_{\alpha^D}) = \pif_D (\h z_{\alpha^D},0) \,.
\ee
We can then sum all the twisted sectors to arrive at
\bea
{\vev{\cO_{0,D} + \tau \cdot \cO_{1,D}}_{g;\n_F} \ov Z_M}=&
{(v+3w)^g \over 2}
\sum_{\alpha^D \in \cS(N_c^D-1,\kay)}
\cO_{0,D} (\h z_{\alpha^D},0)
\cH_{D} (\h z_{\alpha^D},0)^{g-1}
\pif_{D} (\h z_{\alpha^D},0) \\
&+v^g \sum_{\alpha^D \in \cS(N_c^D,\kay)}
\cO_{0,D} (\h z_{\alpha^D})
\cH_{D} (\h z_{\alpha^D})^{g-1}
\pif_{D} (\h z_{\alpha^D})  \,.
\eea
Using the identities of appendix \ref{ap: identities}, we   find:
\bea
{\pif (\h z_\alpha) \over \pif_D (\h z_{\alpha^c},0)}
= {\pif (\h z_\alpha,0) \over \pif_D (\h z_{\alpha^c})}
&= (q_A q_{A,D}^{-1})^{\sum_i \n_i}\, 2^{-2 \sum_i \n_i} \prod_{i \leq j}(-m_i -m_j)^{\n_i + \n_j}\,, \\
{\cH (\h z_\alpha) \over 4\cH_D (\h z_{\alpha^c},0)}
={\cH (\h z_\alpha,0) \over \cH_D (\h z_{\alpha^c})}
&=q_R q_{R,D}^{-1}\, 2^{-4N_c^D+2 \sum_i r_i} e^{i \pi(N_cN_c^D+\nu)} \prod_{i \leq j}(-m_i -m_j)^{1-r_i -r_j} \,.
\eea
The operator map \eq{SO operator map} implies that
\be
\cO_0 (\h z_\alpha) = \cO_{0,D} (\h z_{\alpha^c},0) \,, \qquad
\cO_0 (\h z_\alpha,0) = \cO_{0,D} (\h z_{\alpha^c}) \,,
\ee
$\cO_0$ being symmetric polynomials of the square $\sigma_a^2$ of the Cartan coordinates of $\sigma$. We then arrive at:
\bea
& (q_A q_{A,D}^{-1})^{\sum_i \n_i}(q_R q_{R,D}^{-1})^{g-1} e^{- 2\ln 2 \sum_i \n_i } e^{-((4 \ln 2) N_c^D-2 \ln 2 \sum_i r_i + i \pi (N_cN_c^D + \nu)) (g-1)}\\
&\cdot \vev{\cO_{0,D} + \tau \cdot \cO_{1,D}}_{g;\n_F}  \; =\;
\left( {v+3w \over 4}\right)^{g}
 2 \sum_{\alpha \in \cS(N_c,\kay)}
\cO_{0} (\h z_{\alpha})
\cH (\h z_{\alpha})^{g-1}
\pif (\h z_{\alpha})  \\
&\qquad \qquad \qquad \qquad \qquad \qquad
+v^g \sum_{\alpha \in \cS(N_c-1,\kay)}
\cO_{0} (\h z_{\alpha},0)
\cH (\h z_{\alpha},0)^{g-1}
\pif (\h z_{\alpha},0)  \,.
\eea
If we again take $v=w=1$, we find that the right-hand-side of this equation agrees with equation \eq{orig SO 2N 2k+1}. Thus
\be
\vev{\cO_0 + \text{Pf} (\sigma) \cdot \cO_1 }_{g;\n_F}
= \vev{\cO_{0,D} + \tau \cdot \cO_{1,D}}_{g;\n_F}  \,,
\ee
with
\be
q_{A, D}= e^{-2 \log 2}\,q_A\, , \qquad q_{R,D}=e^{-((4 \ln 2) N_c^D-2 \ln 2 \sum_i r_i + i \pi (N_cN_c^D + \nu))} \,q_R~,
\ee
the relations between contact terms.

\subsubsection{$SO(2N_c+1) \leftrightarrow O_+(2N_c^D)$, $N_f = 2 \kay$, $N_c^D = \kay -N_c$}

\noindent\textbf{Map of vacua :}
$P(z)$ is given by
\be
P(z) = 2 \prod_{\alpha=1}^\kay (z^2 - \h z_\alpha^2)
\ee
The Coulomb branch vacua of the $SO(2N_c+1)$ theory, represented by the vacuum expectation value of the Cartan coordinates $\h \sigma_a$, are given by $N_c$-tuples of roots of $P(z)$ that satisfy the following constraints:
\be
\h \sigma_a \neq \pm m_i ~~ \text{for any }a,~i\,,
\quad
\h \sigma_a \neq 0 ~~ \text{for any }a \,,
\quad
\h \sigma_a \neq \pm \h \sigma_b ~~ \text{for }a \neq b\,,
\ee
up to identifications made under the Weyl group. Then, each vacuum is represented by a tuple of roots:
\be
( \h z_{\alpha_1}, \cdots, \h z_{\alpha_{N_c}} )\,, \quad \alpha \in \cS(N_c,\kay) \,.
\ee
In the dual theory, each vacuum is also represented by a tuple of roots:
\be
( \h z_{\alpha^D_1}, \cdots, \h z_{\alpha^D_{N_c^D}})\,, \quad 
\quad \alpha^D_\ba \in \cS(N_c^D,\kay)
\ee
corresponding to $\alpha^D$. Note that these vacuum expectation values of $\sigma^D$ are not fixed points of the orbifold action, thus having only a single vacuum associated to each expectation value. The duality map is then extremely simple, given by taking complement of a vector $\alpha$ representing vacua of the $SO(2N_c+1)$ theory with respect to $[\kay]$.

\medskip

\noindent\textbf{$A$-twisted correlation functions :} The expectation value of an operator in the $SO(2N_c+1)$ theory is given by
\be
\vev{\cO}_{g;\n_F} = \sum_{\alpha \in \cS(N_c,\kay)} 
\cO(\h z_\alpha) \cH (\h z_\alpha)^{g-1} \pif (\h z_\alpha)
\ee
with
\bea\label{SO 2N+1}
\pif(\h \sigma_a) &= \prod_i \left[ q_A (-m_i) \prod_{a} (m_i^2 -\h \sigma_a^2) \right]^{\n_i} \\
\cH (\h \sigma_a) &= q_R
{\prod_i \left( (-m_i) \prod_a (m_i^2 -\h \sigma^2_a) \right)^{1-r_i} \over 
\prod_{i,a} (\h \sigma_a - m_i)}
\cdot {\prod_a P'(\h \sigma_a) \over \prod_a \h \sigma_a^2
\prod_{a \neq b} (\h \sigma_a^2 - \h \sigma_b^2)} \,.
\eea
Meanwhile, the $O_+(2N^D)$ correlator is obtained by restricting to the untwisted sector only---that is, by setting $v=1$, $w=0$:
\be
\vev{\cO_D}_{g;\n_F}
= {1 \ov 2}\vev{\cO_D}_{g;\n_F,\text{untwisted}}
= Z_M \cdot
\sum_{\alpha^D \in \cS(N_c^D,\kay)}
\cO_D(\h z_{\alpha^D})
\cH_D(\h z_{\alpha^D})
\pif_D(\h z_{\alpha^D}) \,.
\ee
The reason for this particular orbifold prescription should be understood better; we just note that it appears to be necessary to match the correlation functions.
We find that:
\bea
{\pif (\h z_\alpha) \over \pif_D (\h z_{\alpha^c})}
&=(q_A q_{A,D}^{-1})^{\sum_i \n_i} \,2^{-2 \sum_i \n_i} \prod_{i \leq j}(-m_i -m_j)^{\n_i + \n_j} \\
{\cH (\h z_\alpha) \over \cH_D (\h z_{\alpha^c})}
&= q_R q_{R,D}^{-1} \, 2^{-4N^D+2 \sum_i r_i} e^{i \pi(NN^D+\nu+ \kay)} \prod_{i \leq j}(-m_i -m_j)^{1-r_i -r_j} \,,
\eea
for $\pif_D$ and $\cH_D$ given by equation \eq{O 2ND}. By the duality map of operators, we have
$\cO(\h z_\alpha) = \cO_D(\h z_{\alpha^c})$.
We thus arrive at the duality relations:
\be
\vev{\cO}_{g;\n_F}
=  \vev{\cO_D}_{g;\n_F}  \,,
\ee
with the identifications:
\be
q_{A, D}= e^{- 2\ln 2 } \,q_A\, , \qquad q_{R,D}=e^{-((4 \ln 2) N_c^D-2 \ln 2 \sum_i r_i + i \pi (N_cN_c^D + \nu+k))} \,q_R~,
\ee
between contact terms.

\subsubsection{$SO(2N_c+1) \leftrightarrow O_+(2N_c^D+1)$, $N_f = 2 \kay+1$, $N_c^D = \kay -N_c$}

\noindent\textbf{Map of vacua :}
$P(z)$ is given by
\be
P(z) = 2 z \prod_{\alpha=1}^\kay (z^2 - \h z_\alpha^2)
\ee
Every vacuum of the $SO(2N_c+1)$ theory is represented by a tuple of roots:
\be
( \h z_{\alpha_1}, \cdots, \h z_{\alpha_{N_c}} )\,, \quad \alpha \in \cS(N_c,\kay) \,.
\ee
In the dual theory, each vacuum is also represented by a tuple of roots:
\be
( \h z_{\alpha^D_1}, \cdots, \h z_{\alpha^D_{N_c^D}})\,, \quad 
\quad \alpha^D_\ba \in \cS(N_c^D,\kay) \,.
\ee
While these vacuum expectation values of $\sigma^D$ are fixed points of the orbifold action, the orbifold projection in the $O_+(2N^D+1)$ theory is defined so that there is only a single vacuum associated to each expectation value \cite{Hori:2011pd}. The duality map, as before, is given by taking complement of a vector $\alpha$ representing vacua of the $SO(2N_c+1)$ theory with respect to $[\kay]$.

\medskip

\noindent\textbf{$A$-twisted correlation functions :} The expectation value of an operator in the $SO(2N_c+1)$ theory is
\be
\vev{\cO}_{g;\n_F} = \sum_{\alpha \in \cS(N_c,\kay)} 
\cO(\h z_\alpha) \cH (\h z_\alpha)^{g-1} \pif (\h z_\alpha)
\ee
for $\pif$ and $\cH$ defined in \eq{SO 2N+1}. Meanwhile, the $O_+(2N_c^D+1)$ correlator is given by
\be
\vev{\cO_{0,D}}_{g:\n_F} 
= {1 \ov 2} (v+3w)^g \vev{\cO_{0,D}}_{g:\n_F,\text{untwisted}}\,,
\ee
for parameters $v$ and $w$, which depend on the orbifold projection, since the correlation functions in the twisted and untwisted sectors agree. We thus arrive at:
\be
\vev{\cO_D}_{g;\n_F}
= {1 \ov 2}(v + 3w)^g \cdot Z_M \cdot
\sum_{\alpha^D \in \cS(N_c^D,\kay)}
\cO_D(\h z_{\alpha^D})
\cH_D(\h z_{\alpha^D})
\pif_D(\h z_{\alpha^D}) \,,
\ee
with $\pif_D$ and $\cH_D$ given by equation \eq{O 2ND+1}. We find that
\bea
{\pif (\h z_\alpha) \over \pif_D (\h z_{\alpha^c})}
&= (q_A q_{A,D}^{-1})^{\sum_i \n_i} \,2^{-2 \sum_i \n_i} e^{i \pi \sum_i \n_i} \prod_{i \leq j}(-m_i -m_j)^{\n_i + \n_j} \\
{\cH (\h z_\alpha) \over \cH_D (\h z_{\alpha^c})}
&= q_R q_{R,D}^{-1} \,2^{-4N^D -1+2  \sum_i r_i} e^{i \pi(NN^D+N^D+\nu + \sum_i r_i)} \prod_{i \leq j}(-m_i -m_j)^{1-r_i -r_j} \,.
\eea
As always, we have the duality map $\cO(\h z_\alpha) = \cO_D(\h z_{\alpha^c})$ for the operators. We thus arrive at the equality:
\bea
&{\vev{\cO}_{g;\n_F}\ov \vev{\cO_D}_{g;\n_F} }\;&=&\;
 \left({v+3w \ov 2} \right)^{g}\, (q_A q_{A,D}^{-1})^{\sum_i \n_i}(q_R q_{R,D}^{-1})^{g-1}  \\
 &&&\cdot e^{-(2\ln 2 +i\pi)\sum_i \n_i } e^{-((4 \ln 2) N^D-(2 \ln 2+i \pi) \sum_i r_i + i \pi (NN^D +N^D+ \nu)) (g-1)}\,.
\eea
Setting $v =-1$ and $w=1$, we obtain:
\be
\vev{\cO}_{g;\n_F} 
=\vev{\cO_D}_{g;\n_F}  \,,
\ee
with the relations
\be
q_{A, D}= e^{-(2\ln 2 +i\pi)} \,q_A\, , \qquad q_{R,D}=e^{-((4 \ln 2) N^D-(2 \ln 2+i \pi) \sum_i r_i + i \pi (NN^D +N^D+ \nu))}\,q_R~,
\ee
between contact terms.  As before, this particular orbifold prescription is chosen so that the duality relations hold. It would be interesting to understand whether there is a simpler way to fix $v, w$ in each case.

\section{$O_-(N)$ dualities}\label{sec: ON}

Let us now consider the $O_-$ orbifold of theories with $SO(N)$ gauge groups and $N_f$ flavors in the vector representation \cite{Hori:2011pd}. In this particular orbifold projection, the duality maps an $O_-(N)$ theory to an $O_-(N^D)$ theory with
\be
N^D = N_f - N +1 \,.
\ee
As before, the matter content of the $O_-(N)$ theory is given by $N_f$ chiral multiplets $\Phi_i$ ($i=1, \cdots, N_f$) in the vector representation, of $R$-charge $r_i \in \Z$, and we turn on twisted masses and fluxes for the $U(N_f)$  flavor symmetry.
The dual $O_-(N^D)$ theory has $N_f$ chiral fields $\Phi_i^D$ in the vector representation with $R$-charges
\be
r_{D,i} = 1-r_i
\ee
and inverted flavor charges. There are also symmetric mesons $M_{ij}$ in the dual theory and a superpotential
\be
W =  (\Phi_i^D)^t M_{ij}  \Phi^D_j \,.
\ee
The 't Hooft anomalies are again given by \eqref{tHooft SON}.

In the $O_-$ theories, we only concern ourselves with the twisted chiral operators invariant under the Weyl group of the $SO$ group, along with the $\bZ_2$ orbifold group. These are generated by the elementary symmetric polynomials of the Cartan coordinates $\sigma_a$ and $\sigma^D_\ba$. The twisted chiral ring of the dual theories are still summarized by the equation
\be
zP(z) = 2 Q_D(z) Q(z)
\ee
where, as before,
\be
P(z) = \prod_i^{N_f} (z -m_i) +  \prod_i^{N_f} (z +m_i)
\ee
and
\be
Q(z) =\det(z \cdot \mathbf{1} - \sigma) \,, \quad
Q_D(z) =\det(z \cdot \mathbf{1} - \sigma^D) \,.
\ee
There may be twist operators in the twisted chiral spectrum, depending on the orbifold projection. In such cases, the twist operators map into each other:
\be
\tau \leftrightarrow \tau^D \,.
\ee
Let us now examine the dualities and confirm the matching of correlation functions, depending on the parity of $N$ and $N_f$, as in the previous section. Having examined the $SO/O_+$ dualities in detail in the previous section, we will be more concise here.

\subsection{$O_-(2N_c) \leftrightarrow O_-(2N_c^D+1)$, $N_f = 2 \kay$, $N_c^D = \kay -N_c$}

\noindent\textbf{Map of vacua :} $P(z)$ is given by
\be
P(z) = 2 \prod_{\alpha=1}^{\kay} (z^2 - \h z_\alpha^2) \,.
\ee
The vacua of the $O_-(2N_c)$ theory are represented by the tuples
\be
(\h z_{\alpha_1}, \cdots, \h z_{\alpha_{N_c}})\,, \quad
\alpha \in \cS(N_c,\kay) \,.
\ee
These are not fixed points of the orbifold action, and thus only a single vacuum exists for each expectation value. The vacua of the $O_-(2N_c^D+1)$ theory are also represented by the tuples
\be
(\h z^D_{\alpha^D_1}, \cdots, \h z_{\alpha^D_{N_c^D}})\,, \quad
\alpha^D \in \cS(N_c^D,\kay) \,.
\ee
In this case, these vacuum expectation values represent fixed points of the orbifold action. The $O_-$ theory is defined such that only a single vacuum survives the orbifold projection for each of the vacuum expectation values. The map between vacua is summarized by $\alpha \leftrightarrow \alpha^c$, as before.

\medskip

\noindent\textbf{$A$-twisted correlation functions :}
The $A$-twisted correlation function of the $O_-(2N_c)$ theory is given by the untwisted correlation function:
\be
\vev{\cO}_{g;\n_F} =
{1 \ov 2} \vev{\cO}_{g;\n_F, \text{untwisted}}= 
\sum_{\alpha \in \cS(N_c,\kay)}  \cO (\h z_\alpha) \cH(\h z_\alpha)^{g-1} \pif (\h z_\alpha)
\ee
where $\pif$ and $\cH$ are given by \eq{SO 2N}, and the meson contribution $Z_M$ given by \eqref{ZM SO}. In the dual theory, we have:
\bea
\vev{\cO_D}_{g;\n_F}
&= {(v+3w)^g \ov 2} \vev{\cO}_{g;\n_F, \text{untwisted}} \\
&=2^{g-1} \cdot Z_M \cdot
\sum_{\alpha^D \in \cS(N_c^D,\kay)}
\cO_D (\h z_{\alpha^D}) \cH_D (\h z_{\alpha^D})^{g-1} \pif_D (\h z_{\alpha^D}) \,,
\eea
with $\pif_D$ and $\cH_D$ given by \eq{O 2ND+1}. Here we have set $v=-1$ and $w=1$. 
We find that:
\bea
{\pif (\h z_\alpha) \over \pif_D (\h z_{\alpha^c})}
&=  (q_A q_{A,D}^{-1})^{\sum_i \n_i}\, 2^{-2 \sum_i \n_i} e^{i \pi \sum_i \n_i} \prod_{i \leq j}(-m_i -m_j)^{\n_i + \n_j} \,, \\
{\cH (\h z_\alpha) \over \cH_D (\h z_{\alpha^c})}
&=q_R q_{R,D}^{-1}\, 2^{-4N_c^D+2 \sum_i r_i} e^{i \pi(N_cN_c^D+\nu+ \sum_i r_i)} \prod_{i \leq j}(-m_i -m_j)^{1-r_i -r_j}\,,
\eea
and therefore:
\be
\vev{\cO}_{g;\n_F}
= \vev{\cO_D}_{g;\n_F} \,.
\ee
with
\be
q_{A, D}= e^{-(2 \ln 2 +i \pi)}\,q_A\, , \qquad q_{R,D}=e^{(-(\ln 2)(4N_c^D+1)+ i \pi (N_cN_c^D+\nu) + (2 \ln 2 + i \pi) \sum_i r_i)}\,q_R~.
\ee

\subsection{$O_-(2N_c) \leftrightarrow O_-(2N_c^D)$, $N_f = 2 \kay+1$, $N_c^D = \kay -N_c+1$}

\noindent\textbf{Map of vacua :} $P(z)$ is given by
\be
P(z) = 2 z \prod_{\alpha=1}^{\kay} (z^2 - \h z_\alpha^2) \,.
\ee
There are two types of vacua in the $O_-(2N_c)$ theory. The vacua of the first type are represented by the tuples:
\be
(\h z_{\alpha_1}, \cdots, \h z_{\alpha_{N_c}})\,, \quad
\alpha \in \cS(N_c,\kay) \,.
\ee
These are not fixed points of the orbifold action, and there is a single vacuum for each expectation value. The vacua of the second type are represented by the tuples:
\be
(\h z_{\alpha_1}, \cdots, \h z_{\alpha_{N_c-1}},0)\,, \quad
\alpha \in \cS(N_c-1,\kay) \,.
\ee
While these are fixed points of the orbifold action, the orbifold projection leaves a single vacuum for each expectation value.

The vacua of the $O_-(2N_c^D)$ dual theory also come in two varieties. The first are represented by the tuples:
\be
(\h z^D_{\alpha^D_1}, \cdots, \h z_{\alpha^D_{N_c^D-1}},0)\,, \quad
\alpha^D \in \cS(N_c^D-1,\kay) \,.
\ee
The orbifold projection only leaves a single vacuum for each tuple. The vacua of the second type are represented by the tuples:
\be
(\h z^D_{\alpha^D_1}, \cdots, \h z_{\alpha^D_{N_c^D}})\,, \quad
\alpha^D \in \cS(N_c^D,\kay) \,.
\ee
As always, the duality map is obtained by taking the complement of $\alpha$, mapping
\bea\label{sumup Om}
&\alpha \in \cS(N_c,\kay)~ &\Leftrightarrow &\quad~\alpha^c \in \cS(N_c^D-1,\kay)\,, \\
&\alpha \in \cS(N_c-1,\kay)~ &\Leftrightarrow &\quad~\alpha^c \in \cS(N_c^D,\kay) \,.
\eea
In contrast to \eqref{sumup dual rel 532} for the $SO/O_+$ duality,  each vector in \eqref{sumup Om} corresponds to a single vacuum for the $O_-$ duality.

\medskip

\noindent\textbf{$A$-twisted correlation functions :}
The  correlation functions can be straightforwardly computed in every twisted sector, with given $v$ and $w$. For the $O_-(2N_c)$ theory, we find:
\bea
\vev{\cO}_{g;\n_F}
&= v^g \sum_{\alpha \in \cS(N_c,\kay)}
\cO (\h z_{\alpha})
\cH (\h z_{\alpha})^{g-1}
\pif (\h z_{\alpha}) \\
&+ {(v+3w)^g \over 2}
\sum_{\alpha \in \cS(N_c-1,\kay)}
\cO (\h z_{\alpha},0)
\cH (\h z_{\alpha},0)^{g-1}
\pif (\h z_{\alpha},0)\, ,
\eea
with $\pif$ and $\cH$ are given by \eq{SO 2N}. For the $O_-(2N_c^D)$ theory, we obtain:
\bea
{\vev{\cO_D}_{g;\n_F} \ov Z_M}
&= {(v+3w)^g \over 2}
\sum_{\alpha^D \in \cS(N_c^D-1,\kay)}
\cO_{D} (\h z_{\alpha^D},0)
\cH_{D} (\h z_{\alpha^D},0)^{g-1}
\pif_{D} (\h z_{\alpha^D},0) \\
&+v^g \sum_{\alpha^D \in \cS(N_c^D,\kay)}
\cO_{D} (\h z_{\alpha^D})
\cH_{D} (\h z_{\alpha^D})^{g-1}
\pif_{D} (\h z_{\alpha^D})\,,
\eea
where $\pif_D$ and $\cH_D$ are given by \eq{O 2ND}. We have:
\bea
{\pif (\h z_\alpha) \over \pif_D (\h z_{\alpha^c},0)}
= {\pif (\h z_\alpha,0) \over \pif_D (\h z_{\alpha^c})}
&= (q_A q_{A,D}^{-1})^{\sum_i \n_i}\, 2^{-2 \sum_i \n_i} \prod_{i \leq j}(-m_i -m_j)^{\n_i + \n_j} \\
{\cH (\h z_\alpha) \over 4\cH_D (\h z_{\alpha^c},0)}
={\cH (\h z_\alpha,0) \over \cH_D (\h z_{\alpha^c})}
&=q_R q_{R,D}^{-1}\, 2^{-4N_c^D+2 \sum_i r_i} e^{i \pi(N_cN_c^D+\nu)} \prod_{i \leq j}(-m_i -m_j)^{1-r_i -r_j} \,.
\eea
Taking $v=1$, $w=-1$ we find that~\footnote{It would be interesting to understand better this minus sign in the duality relation.}
\be
\vev{\cO}_{g;\n_F}
= -\vev{\cO_D}_{g;\n_F} \,,
\ee
with
\be
q_{A, D}= e^{-2\ln 2}\,q_A\, , \qquad q_{R,D}=e^{(-(\ln 2) (4N_c^D-1) + 2 \ln 2 \sum_i r_i + i \pi (N_cN_c^D + \nu+1))}\,q_R~.
\ee

\subsection{$O_-(2N_c+1) \leftrightarrow O_-(2N_c^D+1)$, $N_f = 2 \kay+1$, $N_c^D = \kay -N_c$}

\noindent\textbf{Map of vacua :}  $P(z)$ is given by
\be
P(z) = 2 \prod_{\alpha=1}^{\kay} (z^2 - \h z_\alpha^2) \,.
\ee
The vacua of the $O_-(2N_c+1)$ theory come in pairs that are represented by the tuples:
\be
(\h z_{\alpha_1}, \cdots, \h z_{\alpha_{N_c}})\,, \quad
\alpha \in \cS(N_c,\kay) \,.
\ee
These are fixed points of the orbifold action, and the orbifold projection keeps two vacua for each  expectation value. The vacua of the $O_-(2N_c^D+1)$ theory also come in pairs represented by the tuples:
\be
(\h z^D_{\alpha^D_1}, \cdots, \h z_{\alpha^D_{N_c^D}})\,, \quad
\alpha^D \in \cS(N_c^D,\kay) \,.
\ee
The two vacua of the $O_-(2N_c+1)$ theory represented by $\alpha \in \cS(N,\kay)$ are mapped to the two vacua in the dual $O_-(2N_c^D+1)$ theory.

\medskip

\noindent\textbf{$A$-twisted correlation functions :}
The correlation function of the $O_-(2N_c+1)$ theory is given by
\be
\vev{\cO}_{g;\n_F}
= {1 \ov 2}{(v+3w)^g}\sum_{\alpha \in \cS(N_c,k)}
\cO(\h z_\alpha) \cH (\h z_\alpha)^{g-1} \pif (\h z_\alpha)
\ee
with $\pif$ and $\cH$ are given by \eq{SO 2N+1}. The $O_+(2N_c^D+1)$ correlator is given by:
\be
\vev{\cO_D}_{g;\n_F}
= {1 \ov 2}(v + 3w)^g \cdot Z_M \cdot
\sum_{\alpha^D \in \cS(N_c^D,\kay)}
\cO_D(\h z_{\alpha^D})
\cH_D(\h z_{\alpha^D})^{g-1}
\pif_D(\h z_{\alpha^D}) \,,
\ee
with $\pif_D$ and $\cH_D$ given by equation \eq{O 2ND+1}. We find that
\bea
{\pif (\h z_\alpha) \over \pif_D (\h z_{\alpha^c})}
&= (q_A q_{A,D}^{-1})^{\sum_i \n_i}\, 2^{-2 \sum_i \n_i} e^{i \pi \sum_i \n_i} \prod_{i \leq j}(-m_i -m_j)^{\n_i + \n_j} \\
{\cH (\h z_\alpha) \over \cH_D (\h z_{\alpha^c})}
&=q_R q_{R,D}^{-1}\, 2^{-4N_c^D -1+2  \sum_i r_i} e^{i \pi(N_cN_c^D+N_c^D+\nu + \sum_i r_i)} \prod_{i \leq j}(-m_i -m_j)^{1-r_i -r_j} \,.
\eea
Setting $v=w=1$, that is, summing over the twisted sectors with equal weight, we find that:
\be
\vev{\cO}_{g;\n_F}
=\vev{\cO_D}_{g;\n_F} \,,
\ee
with the relation:
\be
q_{A, D}=e^{-(2\ln 2+i\pi)}\,q_A\, , \quad q_{R,D}=e^{(-(\ln 2) (4N^D+1) + (2 \ln 2+i\pi) \sum_i r_i + i \pi (NN^D + N^D+\nu))}\,q_R~.
\ee
between contact terms. This completes the proof of the equality of partition functions of Coulomb branch operators across Hori duality.

\section*{Acknowledgements} We would like to thank Ofer Aharony, Stefano Cremonesi, Jonathan Heckman, Heeyeon Kim, Sungjay Lee, Greg Moore,  Wolfger Peelaers, Brian Willett and Alberto Zaffaroni for interesting discussions and comments. CC and DSP gratefully acknowledge support from the Simons Center for Geometry and Physics, Stony Brook University, and from NHETC, Rutgers University, at which some of the research for this paper was performed. DSP would like to thank the physics department at the University of North Carolina at Chapel Hill and  the Korea Institute for Advanced Study for hospitality while this work was being
carried out. NM is supported in part by the ERC Starting Grant 637844-HBQFTNCER and by the INFN. The work of DSP has been supported by DOE grant DOE-SC0010008.


\appendix

\section{Some algebraic identities} \label{ap: identities}

Let us collect some  useful identities, which are used extensively in the main text.
Consider:
\be
P(z) = \prod_{i=1}^{N_f} (z-m_i) +  \prod_{i=1}^{N_f} (z+m_i) \,.
\ee
It is clear that, for any root $\h z$ of $P(z)$,
\be
\prod_i (\h z - m_i) = - \prod_i (\h z + m_i) \,.
\ee
Let us list some basic identities concerning the roots of $P(z)$.
\begin{enumerate}
\item $N_f = 2 \kay$, $P(z) = 2 \prod_{\alpha=1}^\kay (z^2 - \h z_\alpha^2)$.
 \begin{itemize}
 \item $P'(\h z_\beta) = 4 \h z_\beta \prod_{\alpha \neq \beta} (\h z_\beta^2 - \h z_\alpha^2)$.
 \item $\prod_{\alpha} (m_i^2 - \h z_\alpha^2) = {1 \ov 2} \prod_j (m_i + m_j)$
 \item $\prod_{i,\alpha} (m_i^2 - \h z_\alpha^2) = 2^{-2 \kay} \prod_{i, j} (m_i + m_j)$
 \item $\prod_{i,\alpha} (m_i - \h z_\alpha) = (-1)^\kay \prod_{i,\alpha} (m_i + \h z_\alpha) =
  ((-1)^\kay 2^{-2\kay} \prod_{i,j} (m,_i + m_j))^{1/2}$
 \item $\prod_\alpha \h z_\alpha = ((-1)^\ell \prod_i m_i)^{1/2}$
 \item $\prod_{i,\alpha} (m_i -\h z_\alpha) / \prod_\alpha \h z_\alpha = e^{i \nu \pi} \prod_{i < j} (m_i + m_j)$ for an integer $\nu$.
 \item $\prod_\alpha \h z_\alpha \cdot \prod_{i,\alpha} (m_i - \h z_\alpha) = e^{i(\kay + \nu) \pi} 2^{-2\kay} \prod_{i \leq j} (m_i + m_j)$
 \end{itemize}
\item $N_f = 2 \kay+1$, $P(z) = 2 z \prod_{\alpha=1}^\kay (z^2 - \h z_\alpha^2)$.
 \begin{itemize}
 \item $P'(\h z_\beta) = 4 \h z_\beta^2 \prod_{\alpha \neq \beta} (\h z_\beta^2 - \h z_\alpha^2)$.
 \item $P'(0) = 2 (-1)^\kay \prod_{\alpha} \h z_\alpha^2$
 \item $\prod_\alpha(m_i^2-\h z_\alpha^2) = \prod_{j \neq i}(m_i + m_j)$
 \item $\prod_{i,\alpha} (m_i^2 - \h z_\alpha^2) = \prod_{i \neq j} (m_i + m_j)$
 \item $\prod_{i,\alpha} (m_i - \h z_\alpha) = \prod_{i,\alpha} (m_i + \h z_\alpha) =
  e^{i \nu \pi} \prod_{i < j} (m_i + m_j)$ for an integer $\nu$.
 \end{itemize}
\end{enumerate}

\noindent Note that we have introduced an integer $\nu$, defined modulo 2, that determines the phase of certain products. This phase does not depend on the choice of $\h z_\alpha$---taking $\h z_{\alpha_0} \rightarrow -\h z_{\alpha_0}$ for a given index $\alpha_0$ does not alter $\nu$---since
\be
{\prod_i (m_i - \h z_\alpha) \over \h z_\alpha}
= {\prod_i (m_i + \h z_\alpha) \over -\h z_\alpha}
\ee
for $N_f =2\kay$ while 
\be
{\prod_i (m_i - \h z_\alpha) }
= {\prod_i (m_i + \h z_\alpha)}
\ee
for $N_f = 2\kay +1$ for any $\alpha$.

\section{$U(N_c)$ gauge group and Grassmanian duality}\label{app: B}

In this appendix, we present some explicit expressions for the instanton factors of $U(N_c)$ theories. We consider powers of the twisted chiral ring operator:
\be
u_k(\sigma) := \tr(\sigma^k)~.
\ee
These expressions  have interesting relations to invariant quantities on the Grassmanian manifold and some generalisations thereof. Indeed, for $N_a=0$, the $U(N_c)$ theory with $N_f$ fundamentals flows to the $\CN=(2,2)$ supersymmetric NLSM onto the Grassmanian manifold $G(N_c, N_f)$, and the gauge duality reproduces the geometric equivalence:
\be
G(N_c, N_f) \cong G(N_f-N_c, N_f)~,
\ee
which exchanges an hyperplane and its complement.
This geometric interpretation can be generalized to $N_a>0$ \cite{Jia:2014ffa}. More precisely, this interpretation holds only if we take the $R$-charges $r=0$ for the $N_f$ chiral multiplets, $\tilde{r}=1$ for the $N_a$ chiral multiplets, and set to zero the background fluxes, $\n_i=0$.  We will restrict to this setup in the following. We also fix the genus $g=0$. 

In the limit of vanishing twisted masses, the instanton factors give us numbers with an interesting geometric interpretation. For instance, the instanton factors for the $N_a=0$ theory are the Gromov-Witten invariants of the Grassmanian.

\subsection{Instanton level $\mathbf{k}=0$}
We start by considering the instanton factor $\CZ^{[N_c, N_f, N_a]}_{g=0, \mathfrak{m}=0}$ in \eqref{Instanton factor UNc}, with only $u_1(\sigma)$ inserted. It admits a simple expression:
\bea \label{genZ0u1p}
\CZ^{[N_c, N_f, N_a]}_{g=0, \mathfrak{m}=0} (u_1^p (\sigma)) = \sum_{\vec \lambda} \left[{\rm dim} V_{\vec \lambda + ((N_f-N_c)^{N_c})} \right] S_{\vec \lambda}( m_1, \ldots, m_{N_f})~.
\eea
This is a polynomial function of $\vec m=(m_1, \ldots, m_{N_f})$ with 
\bea
m_1 + \ldots + m_{N_f} =0~,
\eea
and is independent of $N_a$.  The notations in the above formula are as follows.
\bi
\item The summation in \eref{genZ0u1p} runs over the partition $\vec \lambda =(\lambda_1 \geq \lambda_2 \geq \cdots \geq \lambda_{N_c} \geq 0)$ of $p-(N_f-N_c)N_c$ into at most $N_c$ parts:
\bea
\lambda_1 + \lambda_2 + \ldots + \lambda_{N_c}=p-(N_f-N_c)N_c~.
\eea
\item The Schur polynomial associated with the partition $\vec \lambda =(\lambda_1 \geq \lambda_2 \geq \cdots \geq \lambda_k \geq 0)$ of an integer into at most $k$ parts is defined as
\bea
S_{\vec \lambda}(m_1, \ldots, m_k) = \frac{|m^{\lambda_i+k-i}|}{|m_j^{k-i}|} = \frac{|m^{\lambda_i+k-i}|}{\prod_{i<j}(m_i -m_j)}~,
\eea
\item For a partition $\vec \mu= (\mu_1 \geq \mu_2 \geq \cdots \geq \mu_\ell \geq 0)$ of $n$, $V_{\vec \mu}$ denotes a representation of the permutation group $S_n$ of $n$ objects.  The dimension of this representation is given by
\bea
\dim V_{\vec \mu} = \frac{n!}{d_1! \cdots d_\ell!} \prod_{i<j} (d_i -d_j)~,
\eea
where
\bea
n= \sum_i \mu_i, \qquad d_i = \lambda_i+\ell -i~.
\eea
\item The notation $(k^m)$ denotes $(\underbrace{k, k, \ldots, k}_{\text{$m$ times}})$.
\ei
\noindent Upon setting $\vec m= \vec 0$, the Schur polynomial becomes
\bea
S_{\vec \lambda} (\vec m= \vec 0) = \begin{cases} 0  & \text{for $\vec \lambda \neq \vec 0$} \\ 1 & \text{for $\vec \lambda =\vec 0$}~.  \end{cases}
\eea
Hence it follows from \eref{genZ0u1p} that
\bea \label{Zq0specialm0}
\CZ^{[N_c, N_f, N_a]}_{g=0, \mathfrak{m}=0} (u_1^p (\sigma)) \Big|_{\vec m =\vec0}&= \left[ \dim V_{((N_f -N_c)^{N_c})}\right]  \delta_{p, (N_f-N_c)N_c}\nn \\
&= \left[ (N_c(N_f-N_c))!\prod_{m=0}^{N_c-1} \frac{m!}{(N_f-N_c+m)!}  \right] \delta_{p, (N_f-N_c)N_c} \nn \\
&= \left[  {\rm deg} \, G(N_f-N_c,N_f) \right] \delta_{p, (N_f-N_c)N_c}~,
\eea
where the quantities in the square brackets are in fact equal to the degree of the Grassmannian $G(N_f-N_c,N_f) \cong G(N_c, N_f)$. 

\paragraph{\it Relation to Schubert calculus.}  The quantity ${\rm deg} \,G(N_f-N_c,N_f)$ in \eref{Zq0specialm0} has a nice geometric interpretation in the context of the Schubert calculus of the Grassmanian.  It is precisely the answer of the following question: given $p =N_c(N_f-N_c)$ general $(N_f-N_c-1)$-planes $L_1, \ldots, L_p$ in $\BP^{N_f-1}$, how many $(N_c-1)$-planes meet all of these $L_i$?  The answer to this question is also equal to the $p$-fold self-intersection number of the Schubert cycle $\sigma_1$ of codimension-one  in $G(N_c, N_f)$.  See also \cite{Chair:1998ex} for a similar exposition. 

\paragraph{\it The operator $u_{n}(\sigma)^p$.}
The instanton factor for the operator $u_n(\sigma)^p$ can also be computed in a similar way. The explicit expression for this is as follows:
\bea
&\CZ_{g=0, {\mathfrak{m}=0}}^{[N_c, N_f, N_a=0]} (u_n(\sigma)^p) \Big |_{\vec m = \vec 0} \nn \\
&= 
\begin{cases}
0 &   \text{if $n \nmid N_c$ and $n \nmid N_f-N_c$} \\
s(N_c, N_f, n) \left[ \left( \frac{N_c(N_f-N_c)}{ n} \right) !\prod_{m=0}^{N_c-1} \frac{\lfloor m/n \rfloor!}{\lfloor (N_f-N_c+m) / n) \rfloor!}  \right] \delta_{p, (N_f-N_c)N_c/n} & \text{if $n | N_c$ or $n |(N_f-N_c)$}~, 
\end{cases}
\eea
where $\lfloor x \rfloor$ denotes the largest integer that is not greater than $x$ and
\bea\label{sgnfn}
s(N_c, N_f, n) = \begin{cases} 
1 & ~ \text{if $N_c$ is odd and ($n| N_c$ or $n|(N_f-N_c)$)} \\ 
(-1)^{N_c(N_f-N_c)/n} & ~ \text{if $N_c$ is even and ($n| N_c$ or $n|(N_f-N_c)$)} ~.
\end{cases}
\eea
Note that for $n=1$, $s(N_c, N_f, 1) =1$ and we reduce to the previous case.

\subsection{Instanton level $\mathbf{k}$}
We now focus on the instanton factor $\CZ^{[N_c, N_f, N_a]}_{g=0, \mathfrak{m}} (u_1^p (\sigma)) \Big|_{\vec m =\vec0}$, with vanishing twisted masses, such that:
\be
\sum_a \mathfrak{m}_a = \mathbf{k}~.
\ee

\subsubsection*{The case of $N_a = 0$}
\paragraph{\it The operator $u_{1}(\sigma)^p$.}
The formula for the partition function in question is
\bea \label{Zku1pm0}
\CZ^{[N_c, N_f, N_a=0]}_{g=0, {\bf k}}(u_{1}(\sigma)^p) \Big|_{\vec m = \vec 0} = \left[ {\rm deg} \; K^{{\bf k}}_{N_f-N_c,N_c} \right]  \delta_{p, (N_f-N_c)N_c + {\bf k} N_f}
\eea
where $K^{\bf k}_{N_f-N_c, N_c}$ is the space of rational curves of degree ${\bf k}$ on the Grassmanian variety $G(N_c, N_f) \cong G(N_f-N_c, N_f)$.   There is an isomorphism
\bea \label{Kduality}
K^{\bf k}_{N_f-N_c, N_c} \cong K^{\bf k}_{N_c, N_f-N_c}
\eea
Note that for ${\bf k} =0$, this space can be identified with the Grassmannian itself:
\bea
K^{0}_{N_f-N_c, N_c} \cong G(N_c,N_f)~.
\eea
The degree of this space was computed in \cite{RRW}:
\bea
&{\rm deg} \; K^{{\bf k}}_{N_f-N_c,N_c} \nn \\
&= (-1)^{{\bf k}(N_f-N_c+1)} \left[ (N_f-N_c)N_c + {\bf k} N_f\right]! \times \nn \\
&\qquad \sum_{n_1+\ldots+n_{N_f-N_c} = {\bf k}} \frac{ \prod_{1 \leq k<j\leq N_f-N_c}\left[ (j-k)+(n_j-n_k)N_f \right]}{\prod_{j=1}^{N_f-N_c} (N_c +j +n_j N_f -1)!}  \\
&= (-1)^{{\bf k}(N_c+1)} \left[ (N_f-N_c)N_c + {\bf k} N_f\right]! \times \nn \\
&\qquad \sum_{n_1+\ldots+n_{N_c} = {\bf k}} \frac{ \prod_{1 \leq k<j\leq N_c}\left[ (j-k)+(n_j-n_k)N_f \right]}{\prod_{j=1}^{N_c} (N_f-N_c +j +n_j N_f -1)!} \\
&=  (-1)^{{\bf k}(N_c+1)} (-1)^{N_c(N_c-1)/2}  (N_c(N_f-N_c) + {\bf k} N_f)! \times  \nn \\
& \qquad \sum_{n_1+\ldots+n_{N_c} = {\bf k}} \; \sum_{\sigma \in S_{N_c} } \; \prod_{j=1}^{N_c} \frac{1}{\left( N_f-2N_c-1+j +\sigma(j)+n_j N_f \right)!}~. \label{degKk}
\eea
Due to the duality \eref{Kduality}, it follows that
\be
\CZ^{[N_c, N_f, N_a=0]}_{g=0, {\bf k}}(u_{1}(\sigma)^p) \Big|_{\vec m = \vec 0} = \CZ^{[N_c, N_f-N_c, N_a=0]}_{g=0, {\bf k}}(u_{1}(\sigma)^p) \Big|_{\vec m = \vec 0}~.
\ee
This equality is in agreement with the GLSM duality.

\paragraph{\it The special case of $(N_c, N_f, N_a) =(2, 5, 0)$.}
In the special case of $N_c=2$ and $N_f=5$, it is interesting to point out that the degree of $K^{\bf k}_{3,2}$ is a Fibonacci sequence
\bea
\deg K^{\bf k}_{3,2} = F(5{\bf k} +5)~, \label{degK32k}
\eea
where
\bea
F(m) = \frac{1}{\sqrt{5}} \left[ \left( \frac{1+\sqrt{5}}{2} \right)^{m}-  \left( \frac{1-\sqrt{5}}{2} \right)^{m}  \right]~.
\eea 
This model was also studied in detail in \cite{Intriligator:1991an} (see also Eq. (26) of \cite{Chair:1998ex}).
The instanton factor is given by:
\bea
& \CZ_{g=0, {\bf k}}^{N_c =2, N_f=5, N_a=0} (u_n(\sigma)^p) \Big |_{\vec m = \vec 0}  \nn \\
&= \begin{cases} 
0 & \quad \text{if $n \nmid (6+5{\bf k})$} \\
\mathfrak{s}(n, {\bf k}) \; F\left(\frac{6+5{\bf k}}{n}+ a(n) \right) \delta_{p,(6+5{\bf k})/n} & \quad \text{if $n | (6+5{\bf k})$}~,
\end{cases}
\eea
where $F(m)$ denotes the Fibonacci number
\bea
F(m) = \frac{1}{\sqrt{5}} \left[ \left( \frac{1+\sqrt{5}}{2} \right)^{m}-  \left( \frac{1-\sqrt{5}}{2} \right)^{m}  \right]~,
\eea
the function $a(n)$ is given by
\bea
a(n) = \begin{cases} -1 & ~\text{if $n \equiv \pm1~ (\mathrm{mod}\; 5)$}\\ 1 & ~\text{if $n \equiv \pm 2~ (\mathrm{mod}\; 5)$} \\ 0  & ~\text{if $n \equiv 0~ (\mathrm{mod}\; 5)$} \end{cases}~,
\eea
and the function $\mathfrak{s}(n, {\bf k})$ is given by
\bea
\mathfrak{s}(n, {\bf k}) = \begin{cases} 
1  & ~\text{if $n$ is odd, $n | (6+5{\bf k})$ and $n \nmid 6$} \\
(-1)^{6/n}  & ~\text{if $n$ is odd, $n | (6+5{\bf k})$ and $n |6$} \\
(-1)^{(6+5 {\bf k})/n}  & ~\text{if $n$ is even and $n | (6+5{\bf k})$} 
\end{cases}
\eea

\subsubsection*{General value of $N_a$}
The function $\CZ^{[N_c,N_f,N_a]}_{g=0, {\bf k }}(u_{1}(\sigma)^p)$ can be written as 
\bea  \label{ZNcNfNak}
\CZ^{[N_c,N_f,N_a]}_{g=0,{\bf k}}(u_{1}(\sigma)^p) \Big |_{\vec m=0, \tilde{\vec m} =0} = \left[ {\rm deg} \; \CM^{{\bf k}}_{N_c, N_f, N_a} \right]  \delta_{p, (N_f-N_c)N_c + {\bf k} (N_f-N_a)}
\eea
where
\bea
 &{\rm deg} \; \CM^{{\bf k}}_{N_c, N_f, N_a} \nn \\
&= (-1)^{{\bf k}(N_c+N_a+1)} (-1)^{N_c(N_c-1)/2}  (N_c(N_f-N_c) + {\bf k} (N_f-N_a))! \times  \nn \\
& \quad \sum_{n_1+\ldots+n_{N_c} = {\bf k}} \; \sum_{\sigma \in S_{N_c} } \; \prod_{j=1}^{N_c} \frac{1}{\left[ N_f-2N_c-1+j +\sigma(j)+n_j (N_f-N_a) \right]!}~. \label{degMgen}
\eea

\paragraph{\it The special case of $N_f = N_a+1$.}
In this case, formula \eref{ZNcNfNak} reduces to
\bea 
&\CZ^{[N_c, N_f, N_f-1]}_{{\bf k}}(u_{1}(\sigma)^p) \Big |_{\vec m=0, \tilde{\vec m} =0}  \nn \\
&=(-1)^{{\bf k}(N_f-N_c)} \; \left[ {\rm dim} \; V_{\left({\bf k} +(N_f-N_c), \; (N_f-N_c)^{N_c-1} \right)} \right] \delta_{p, N_c(N_f-N_c)+{\bf k} } ~,
\eea
where
\bea
& {\rm dim} \; V_{\left({\bf k} +(N_f-N_c), \; (N_f-N_c)^{N_c-1} \right)} \nn \\
&= \frac{({\bf k} +N_c(N_f-N_c))!}{({\bf k}+N_f-1)!}  \prod_{m=0}^{N_c-1} \frac{ (m-1)!(m+{\bf k})}{(N_f-N_c+m-1)! }~.
\eea

\subsection{The case $N_f=N_a$}
We find that the resummed expectation value of $u_1^{p}$ can be written as follows:
\bea
\langle u_1^{p} \rangle \Big|_{\vec m=  \tilde{\vec m} =0} & = \sum_{{\bf k} =0}^\infty {\mathcal Z}^{[N_c,N_f=N_a]}_{g=0, {\bf k}} ( u_1^{p}) q^{\bf k}  \nn \\
& = \begin{cases}  
\frac{{\rm deg} \; G(N_f-N_c,N_c)}{\left[ 1+(-1)^{N_f-N_c} q \right]^{N_c}} & \quad \text{if $p= N_c(N_f-N_c)$} \\
0 & \quad \text{otherwise}~,
\end{cases}
\eea
where ${\rm deg} \; G(N_f-N_c,N_c)$ is the degree of Grassmannian $G(N_f-N_c,N_c)$, whose explicit expression is given above. 
Similarly,  the (resummed) expectation value of $u_n^{p}$ is given by
\bea
&\langle u_n(\sigma)^{p} \rangle \Big|_{\vec m=  \tilde{\vec m} =0}  \nn \\
&= \sum_{{\bf k} =0}^\infty {\mathcal Z}^{(N_c,N_f=N_a)}_{g=0, {\bf k}} ( u_1^{p}) q^{\bf k} \nn \\  
&=
\begin{cases} 
0 &   \text{if $n \nmid N_c$ and $n \nmid N_f-N_c$} \\
s(N_c, N_f, n) \left[ \left( \frac{N_c(N_f-N_c)}{ n} \right) !\prod_{m=0}^{N_c-1} \frac{\lfloor m/n \rfloor!}{\lfloor (N_f-N_c+m) / n) \rfloor!}  \right]  \times \nn \\
\qquad {\left[ 1+(-1)^{N_f-N_c} q \right]^{-N_c}} \; \delta_{p, N_c(N_f-N_c) -(n-1)N'} &  \text{if $n | N_c$ or $n | N_f-N_c$}~.
\end{cases}
\eea
where $s(N_c, N_f, n)$ is given by \eref{sgnfn} and
\bea
N' = \begin{cases} N_f- N_c & \quad \text{if $n|N_c$}  \\ N_c & \quad \text{if $n|(N_f-N_c)$} \end{cases}~.
\eea

\subsubsection*{Duality}
We find that
\bea
\left[ \sum_{{\bf k} =0}^\infty {\mathcal Z}^{[N_c,N_f=N_a]}_{g=0, {\bf k}} ( u_n (\sigma)^{p}) q^{\bf k} \right]  = f(q) \left[ \sum_{{\bf k} =0}^\infty {\mathcal Z}^{D, [N_f-N_c,N_a=N_f]}_{g=0, -{\bf k}} ( -u_n(\sigma)^{p}) q_D^{\bf k} \right]~,
\eea
where
\bea
f(q) = \left( 1+(-1)^{N_f-N_c}q \right)^{N_f-2N_c}~, \qquad q_D = (-1)^{N_f} q^{-1}~.
\eea
Note that both sides of the equality are non-zero if and only if 
\bea p = N_c(N_f-N_c) -(n-1)N'~. \eea

%
\bibliographystyle{JHEP}
\bibliography{bib2d}{}

\end{document}